\documentclass[iop,revtex4-1]{emulateapj}

\usepackage{natbib}
\usepackage{hyperref}

\usepackage{amsmath}
\usepackage{xspace}
\usepackage{lineno}

\newcommand{\code}[1]{\textsc{#1}\xspace}
\newcommand{\degree}{\ensuremath{{}^{\circ}}\xspace}
\newcommand{\solar}{\ensuremath{{}_{\odot}}\xspace}

\shorttitle{Eridanus-Phoenix Overdensity}
\shortauthors{Li et al.}

\begin{document}

\title{Discovery of a Stellar Overdensity in Eridanus-Phoenix in the Dark Energy Survey}

\author{
T.~S.~Li\altaffilmark{1},
E.~Balbinot\altaffilmark{2},
N.~Mondrik\altaffilmark{1,3},
J.~L.~Marshall\altaffilmark{1},
B.~Yanny\altaffilmark{4},
K.~Bechtol\altaffilmark{5},
A.~Drlica-Wagner\altaffilmark{4},
D.~Oscar\altaffilmark{6,7},
B.~Santiago\altaffilmark{6,7},
J.~D.~Simon\altaffilmark{8},
A.~K.~Vivas\altaffilmark{9},
A.~R.~Walker\altaffilmark{9},
M.~Y.~Wang\altaffilmark{1},
T. M. C.~Abbott\altaffilmark{9},
F.~B.~Abdalla\altaffilmark{10},
A.~Benoit-L{\'e}vy\altaffilmark{10},
G.~M.~Bernstein\altaffilmark{11},
E.~Bertin\altaffilmark{12,13},
D.~Brooks\altaffilmark{10},
D.~L.~Burke\altaffilmark{14,15},
A.~Carnero~Rosell\altaffilmark{7,16},
M.~Carrasco~Kind\altaffilmark{17,18},
J.~Carretero\altaffilmark{19,20},
L.~N.~da Costa\altaffilmark{7,16},
D.~L.~DePoy\altaffilmark{1},
S.~Desai\altaffilmark{21,22},
H.~T.~Diehl\altaffilmark{4},
P.~Doel\altaffilmark{10},
J.~Estrada\altaffilmark{4},
D.~A.~Finley\altaffilmark{4},
B.~Flaugher\altaffilmark{4},
J.~Frieman\altaffilmark{4,5},
D.~Gruen\altaffilmark{23,24},
R.~A.~Gruendl\altaffilmark{17,18},
G.~Gutierrez\altaffilmark{4},
K.~Honscheid\altaffilmark{25,26},
D.~J.~James\altaffilmark{9},
K.~Kuehn\altaffilmark{27},
N.~Kuropatkin\altaffilmark{4},
O.~Lahav\altaffilmark{10},
M.~A.~G.~Maia\altaffilmark{7,16},
M.~March\altaffilmark{11},
P.~Martini\altaffilmark{25,28},
R.~Ogando\altaffilmark{7,16},
A.~A.~Plazas\altaffilmark{29},
K.~Reil\altaffilmark{14,15}, 
A.~K.~Romer\altaffilmark{30},
A.~Roodman\altaffilmark{14,15},
E.~Sanchez\altaffilmark{31},
V.~Scarpine\altaffilmark{4},
M.~Schubnell\altaffilmark{32},
I.~Sevilla-Noarbe\altaffilmark{31,17},
R.~C.~Smith\altaffilmark{9},
M.~Soares-Santos\altaffilmark{4},
F.~Sobreira\altaffilmark{4,7},
E.~Suchyta\altaffilmark{25,26},
M.~E.~C.~Swanson\altaffilmark{18},
G.~Tarle\altaffilmark{32},
D.~Tucker\altaffilmark{4},
Y.~Zhang\altaffilmark{32}
\\ \vspace{0.2cm} (The DES Collaboration) \\
}
 
\altaffiltext{1}{George P. and Cynthia Woods Mitchell Institute for Fundamental Physics and Astronomy, and Department of Physics and Astronomy, Texas A\&M University, College Station, TX 77843,  USA}\email{email: sazabi@neo.tamu.edu}
\altaffiltext{2}{Department of Physics, University of Surrey, Guildford GU2 7XH, UK}
\altaffiltext{3}{Department of Physics, Harvard University, Cambridge, MA 02138, USA}
\altaffiltext{4}{Fermi National Accelerator Laboratory, P. O. Box 500, Batavia, IL 60510, USA}
\altaffiltext{5}{Kavli Institute for Cosmological Physics, University of Chicago, Chicago, IL 60637, USA}
\altaffiltext{6}{Instituto de F\'\i sica, UFRGS, Caixa Postal 15051, Porto Alegre, RS - 91501-970, Brazil}
\altaffiltext{7}{Laborat\'orio Interinstitucional de e-Astronomia - LIneA, Rua Gal. Jos\'e Cristino 77, Rio de Janeiro, RJ - 20921-400, Brazil}
\altaffiltext{8}{Carnegie Observatories, 813 Santa Barbara St., Pasadena, CA 91101, USA}
\altaffiltext{9}{Cerro Tololo Inter-American Observatory, National Optical Astronomy Observatory, Casilla 603, La Serena, Chile}
\altaffiltext{10}{Department of Physics \& Astronomy, University College London, Gower Street, London, WC1E 6BT, UK}
\altaffiltext{11}{Department of Physics and Astronomy, University of Pennsylvania, Philadelphia, PA 19104, USA}
\altaffiltext{12}{CNRS, UMR 7095, Institut d'Astrophysique de Paris, F-75014, Paris, France}
\altaffiltext{13}{Sorbonne Universit\'es, UPMC Univ Paris 06, UMR 7095, Institut d'Astrophysique de Paris, F-75014, Paris, France}
\altaffiltext{14}{Kavli Institute for Particle Astrophysics \& Cosmology, P. O. Box 2450, Stanford University, Stanford, CA 94305, USA}
\altaffiltext{15}{SLAC National Accelerator Laboratory, Menlo Park, CA 94025, USA}
\altaffiltext{16}{Observat\'orio Nacional, Rua Gal. Jos\'e Cristino 77, Rio de Janeiro, RJ - 20921-400, Brazil}
\altaffiltext{17}{Department of Astronomy, University of Illinois, 1002 W. Green Street, Urbana, IL 61801, USA}
\altaffiltext{18}{National Center for Supercomputing Applications, 1205 West Clark St., Urbana, IL 61801, USA}
\altaffiltext{19}{Institut de Ci\`encies de l'Espai, IEEC-CSIC, Campus UAB, Carrer de Can Magrans, s/n,  08193 Bellaterra, Barcelona, Spain}
\altaffiltext{20}{Institut de F\'{\i}sica d'Altes Energies, Universitat Aut\`onoma de Barcelona, E-08193 Bellaterra, Barcelona, Spain}
\altaffiltext{21}{Excellence Cluster Universe, Boltzmannstr.\ 2, 85748 Garching, Germany}
\altaffiltext{22}{Faculty of Physics, Ludwig-Maximilians University, Scheinerstr. 1, 81679 Munich, Germany}
\altaffiltext{23}{Max Planck Institute for Extraterrestrial Physics, Giessenbachstrasse, 85748 Garching, Germany}
\altaffiltext{24}{Universit\"ats-Sternwarte, Fakult\"at f\"ur Physik, Ludwig-Maximilians Universit\"at M\"unchen, Scheinerstr. 1, 81679 M\"unchen, Germany}
\altaffiltext{25}{Center for Cosmology and Astro-Particle Physics, The Ohio State University, Columbus, OH 43210, USA}
\altaffiltext{26}{Department of Physics, The Ohio State University, Columbus, OH 43210, USA}
\altaffiltext{27}{Australian Astronomical Observatory, North Ryde, NSW 2113, Australia}
\altaffiltext{28}{Department of Astronomy, The Ohio State University, Columbus, OH 43210, USA}
\altaffiltext{29}{Jet Propulsion Laboratory, California Institute of Technology, 4800 Oak Grove Dr., Pasadena, CA 91109, USA}
\altaffiltext{30}{Department of Physics and Astronomy, Pevensey Building, University of Sussex, Brighton, BN1 9QH, UK}
\altaffiltext{31}{Centro de Investigaciones Energ\'eticas, Medioambientales y Tecnol\'ogicas (CIEMAT), Madrid, Spain}
\altaffiltext{32}{Department of Physics, University of Michigan, Ann Arbor, MI 48109, USA}

\begin{abstract}
We report the discovery of an excess of main sequence turn-off stars in the direction of the constellations of Eridanus and Phoenix from the first year data of the Dark Energy Survey (DES). The Eridanus-Phoenix (EriPhe) overdensity is centered around  $l\sim 285\degree$ and $b\sim -60\degree$ and spans at least 30\degree in longitude and 10\degree in latitude. The Poisson significance of the detection is at least $9\sigma$. The stellar population in the overdense region is similar in brightness and color to that of the nearby globular cluster NGC 1261, indicating that the heliocentric distance of EriPhe is about $d\sim16$~kpc. The extent of EriPhe in projection is therefore at least $\sim4$~kpc by $\sim3$~kpc. On the sky, this overdensity is located between NGC 1261 and a new stellar stream discovered by DES at a similar heliocentric distance, the so-called Phoenix Stream. Given their similar distance and proximity to each other, it is possible that these three structures may be kinematically associated. Alternatively, the EriPhe overdensity is morphologically similar to the Virgo overdensity and the Hercules-Aquila cloud, which also lie at a similar Galactocentric distance. These three overdensities lie along a polar plane separated by $\sim$~120\degree and may share a common origin. Spectroscopic follow-up observations of the stars in EriPhe are required to fully understand the nature of this overdensity.

\end{abstract}


\section{INTRODUCTION}
\label{sec:intro}

Great progress has been made towards a better understanding of the formation mechanisms of the Milky Way and especially its stellar halo in recent years. This is due in large part to the deep, wide-field imaging surveys such as the Two Micron All Sky Survey~\citep[2MASS;][]{Skrutskie2006} and the Sloan Digital Sky Survey~\citep[SDSS;][]{York2000}, which   have enabled large-scale studies of the Milky Way Galaxy to its outermost radii, and have resulted in the discovery of a variety of new and interesting substructures that lie within the Milky Way's halo \citep{Ivezic2012}. Our understanding of the formation of the Galactic halo has therefore evolved from a simple Eggen, Lynden-Bell \& Sandage~\citep[ELS; ][]{Eggen1962} monolithic collapse model to a much more complex and dynamic structure that is still being shaped by the merging of neighboring smaller galaxies~\citep{Searle1978}. This merger process is predicted by hierarchical $\Lambda$CDM models of galaxy formation~\citep{Steinmetz2002, Bullock2005, Font2011}. 

In addition to classical globular clusters~\citep{Harris1996, Harris2010} and classical dwarf galaxies~\citep{Mateo1998}, these modern surveys have also discovered many low luminosity globular clusters~\citep{Koposov2007, Balbinot2013}, a large number of ultra faint dwarf galaxies \citep{Willman2005a, Willman2005b, Belokurov2006a,  Zucker2006, Belokurov2007, Walsh2007},  and stellar streams,  which are thought to have originated from the tidal disruption of either dwarf galaxies~\citep[see, e.g., the Sagittarius stream; ][]{Majewski2003, Belokurov2006b} or globular clusters~\citep[see, e.g., the Palomar 5 tidal tails;][]{Odenkirchen2001}. 

However, there are other structures that cannot be classified by one of these familiar designations. Modern wide field deep imaging surveys have enabled for the first time the discovery of large scale highly diffuse halo structures that can cover up to several hundred square degrees on the sky. Examples of these recently identified extended stellar distributions include the Virgo overdensity \citep{Juric2008} and the Hercules-Aquila cloud \citep{Belokurov2007}, both located about 20 kpc from the Sun, and the Pisces overdensity~\citep{Sesar2007} at a distance of about 85 kpc.  

These overdensities have generally been identified by noting excesses of a particular stellar tracer in a restricted area of the sky.  For example, excess halo stars were first found in the Virgo constellation \citep{Vivas2001}, $(l,b) = (285\degree, 60\degree)$, by using RR Lyrae stars as tracers.  This structure is known as the Virgo stellar stream (VSS) and lies about 20 kpc from the Sun \citep{Duffau2006}.  In a similar location in the sky,  \citet{Juric2008}, working with a large sample of Milky Way disk and halo stars from SDSS, identified a very extended overdensity that stood out from smooth models of the Galaxy's thin/thick disks and halo at a heliocentric distance of $\sim$6--20~kpc. This structure is dubbed the Virgo overdensity (VOD) and appears to encompass an area as large as 1,000 deg$^2$. \citet{Bonaca2012} used a later data release from SDSS and estimated that the VOD spans at least 2,000 deg$^2$, with the true extent likely closer to 3,000 deg$^2$. There is likely a connection between the VSS and the VOD; however, the exact nature of that connection is still being debated as is the full orbit of the Virgo structures \citep{Carlin2012}.

The Hercules-Aquila (HerAq) cloud is another nearby, extended structure, covering a few hundred square degrees in the constellations of Hercules and Aquila.  In Galactic coordinates it is located near $l=40\degree$ and spans from $b = +40\degree$ to $-40\degree$ in latitude.  Its distance, inferred from the brightness and colors of the stars, is about 10-20~kpc from the Sun.  \citet{Simion2014} also found a strong excess of RR Lyrae stars in a similar part of the sky, with a distance peaking at 18~kpc from the Sun.

The other similar halo structure currently known is the Pisces Overdensity (POD), which was identified at a distance of $\sim85$ kpc in Pisces using RR Lyrae stars~\citep{Sesar2007, Watkins2009}. Spectroscopic follow-up observations on the RR Lyrae stars confirmed that it is a genuine structure and suggested that it may be composed of two distinct kinematic groups~\citep{Kollmeier2009,Sesar2010}. Analysis of M-giant candidates also showed an overdensity in approximately the same direction \citep{Sharma2010}. Furthermore, \citet{Nie2015} used the blue horizontal branch stars as tracers to determine its spatial extent, finding that POD may be part of a stream with a clear distance gradient.

These structures share a common diffuse cloud-like morphology with large extended structure in the sky. It is still unclear what the origin of these structures might be; they may be the remnants of dwarf galaxies that have been severely disrupted by interaction with the Milky Way's tidal field. It is also possible that they are stellar streams in the later stages of disruption, or streams which are viewed at certain points in their orbit (apogalacticon, perigalacticon) where the tidal dispersal mechanism is particularly effective  \citep{Johnston2008,Johnston2012}.   

There are also two prominent low-latitude, extended overdensities known, the Monoceros Ring structure \citep{Newberg2002,Yanny2003} and the Triangulum Andromeda overdensity \citep[TriAnd; ][]{Rocha2004}, one or both of which may be classified with the above extended overdensities. The origins of these two overdensities are still under debate. While many studies argue that the Monoceros Ring and TriAnd are the remnants of past accretion events~\citep{Crane2003, Martin2004, Penarrubia2005, Sollima2011, Slater2014}, recent works suggest the possibility that the Monoceros Ring and TriAnd may be the result of a strong oscillation in the outer disk which throws thin disk and thick disk stars to large scale heights~\citep{Xu2015, PriceWhelan2015}.

\ 

We report here the discovery of an extended, diffuse stellar overdensity in the direction of constellations Eridanus and Phoenix using imaging from the first year of the Dark Energy Survey. We refer to this structure as the Eridanus-Phoenix (EriPhe) overdensity in this paper. Much like the structures described above, EriPhe covers an area of at least 100 deg$^2$ and has a  heliocentric distance of $\sim$16 kpc.

We structure the paper as follows: in Section 2, we describe our data sample and our search for extended overdensities using main sequence turn-off (MSTO) stars. We then perform a series of checks to confirm that the overdensity is genuine. We also estimate the surface brightness and total luminosity of EriPhe. In Section 3, we discuss the relation of this new stellar overdensity candidate to the other known overdensity structures.  We also search the vicinity of EriPhe and note a possible association between this structure, the globular cluster NGC 1261, and a newly discovered stellar stream~\citep[Phoenix Stream, ][]{Balbinotip}. We conclude in Section 4.

\section{Data and Analysis}
\label{sec:data}

\subsection{The Dark Energy Survey}

The Dark Energy Survey \citep[DES,][]{Abbott2005} is a wide-field optical imaging survey in the $grizY$ bands performed with the Dark Energy Camera (DECam) \citep{flaugher_2015_decam}, which is installed at the prime focus of the 4-meter Blanco telescope at Cerro Tololo Inter-American Observatory. The DECam focal plane is comprised of  74 CCDs: 62 2k$\times$4k CCDs dedicated to science imaging and 12 2k$\times$2k CCDs for guiding, focus, and alignment. DECam has a hexagonal 2.2 deg wide field-of-view and a central pixel scale of 0.263 arcseconds. The full survey is scheduled for 525 nights distributed over five years from 2013 August to 2018 February. This five year survey will eventually cover 5,000 deg$^2$ in the Southern Galactic Cap to depths of $g = 24.6$, $r = 24.1$, $i = 24.4$, $z = 23.8$, and $Y = 21.3$ mag with 10 sigma detection for point sources~\citep{Rossetto2011}. The first year of imaging data~\citep{Diehl2014} have been processed and the calibrated images are now public\footnote{http://www.portal-nvo.noao.edu/}.  DES has internally constructed catalogs of sources from this first year of imaging data, which we refer to as the Year 1 Annual 1 (Y1A1) release, and we use the catalogs from the internal release to search for coherent stellar structures in the halo.   

\subsection{Sample}
\label{sec:sample}

The Y1A1 release consists of $\sim$ 12,000 science exposures taken between 2013 August~15 and 2014 February~9 in the wide survey area. Most of the Y1A1 footprint is covered by 2 to 4 overlapping exposures. Each exposure is 90 s in $griz$ and 45 s in $Y$. The Y1A1 data release covers $\sim$1,800 deg$^2$, including $\sim$200 deg$^2$ overlapping with the Stripe 82 region of SDSS and a contiguous region of $\sim$1,600 deg$^2$ overlapping the South Pole Telescope (SPT) footprint \citep{Carlstrom2011}.  This paper considers the data from the SPT footprint.

The science exposures are processed by the DES data management (DESDM) infrastructure \citep{Gruendlip}. The pipeline consists of image detrending, astrometric calibration, nightly photometric calibration, global calibration, image coaddition, and object catalog creation. The \code{SExtractor} toolkit is used to create object catalogs from the processed and coadded images \citep{Bertin1996, Bertin2011}. The Y1A1 data release contains a catalog of $\sim$ 131 million unique objects detected in the coadded images. For a more detailed description of the DESDM image processing pipeline and the Y1A1 release, we refer to \citet{Sevilla2011}, \citet{Mohr2012}, \citet{Balbinot2015}.

We selected our stellar sample from the Y1A1 coadd object catalog. The objects in the sample are classified using $spread\_model$, a star-galaxy indicator, provided by \code{SExtractor}. The  $spread\_model$ variable is a normalized linear discriminant between the best-fit model of local point spread function (PSF) and a more spatially extended model. A good criterion to select stars is $|spread\_model| < 0.003$~\citep{Desai2012}.  To avoid issues arising from fitting the PSF across variable-depth coadded images, we utilized the weighted-average (wavg) of the $spread\_model$ measurements from the single-epoch exposures. Here we use the weighted-average of the $spread\_model$ measurements in $i$-band $|wavg\_spread\_model\_i|<0.003$ for star selection. \citet{Bechtol2015} shows that with this star selection criteria, the stellar completeness is $>$ 90\% down to magnitude $g \sim 22$, at which point it drops to $\sim$ 50\% by $g \sim 23$. We selected stars with $16 < g < 23$, as fainter sources ($g > 23$) have an increased galaxy contamination rate, and brighter sources ($g < 16$) are typically saturated in DECam exposures. We also apply two additional criteria based on the output from SExtractor, $ flags\_{\{g, r, i\}} < 4$ \footnote{$ flags < 4$ implies that object may have bright neighbors ($flags=1$) and/or might be deblended  ($flags=2$).} to ensure sample quality and $magerr\_psf\_{\{g, r, i\}} < 1$\footnote{$magerr\_psf\_{\{g, r, i\}}< 1$  implies the photmetric errors in PSF magnitude are less than 1 mag in $g-$, $r-$ and $i-$band.} to discard objects with large photometric errors.  The typical statistical photometric errors for the sample are $\sim$0.02 for $g<21.5$ in the $g$-band and increase to $\sim0.05$ at the faint limit ($g\sim23$).

Finally the sample is corrected for Galactic extinction using dust maps from \citet{Schlegel1998} and the recalibrated scaling relation from \citet{Schlafly2011} according to a~\citet{Fitzpatrick1999} reddening law.

\subsection{Overdensity in MSTO stars}
\label{sec:MSTO}
MSTO stars have been used extensively in SDSS as a tracer to map the halo structure and probe new substructures within 30~kpc~\citep[see, e.g.,][]{Newberg2002, Juric2008, Newberg2010}. Here, we use our sample to select halo MSTO stars by including only those stars having $0.2 < (g-r)_0 < 0.4$, where the subscript 0 denotes the dereddened magnitude.  We then study the density map of MSTO stars in different magnitude bins (i.e. different distances) to search for overdensities via visual inspection.

\begin{figure*}[th!]
\epsscale{1} 
\plotone{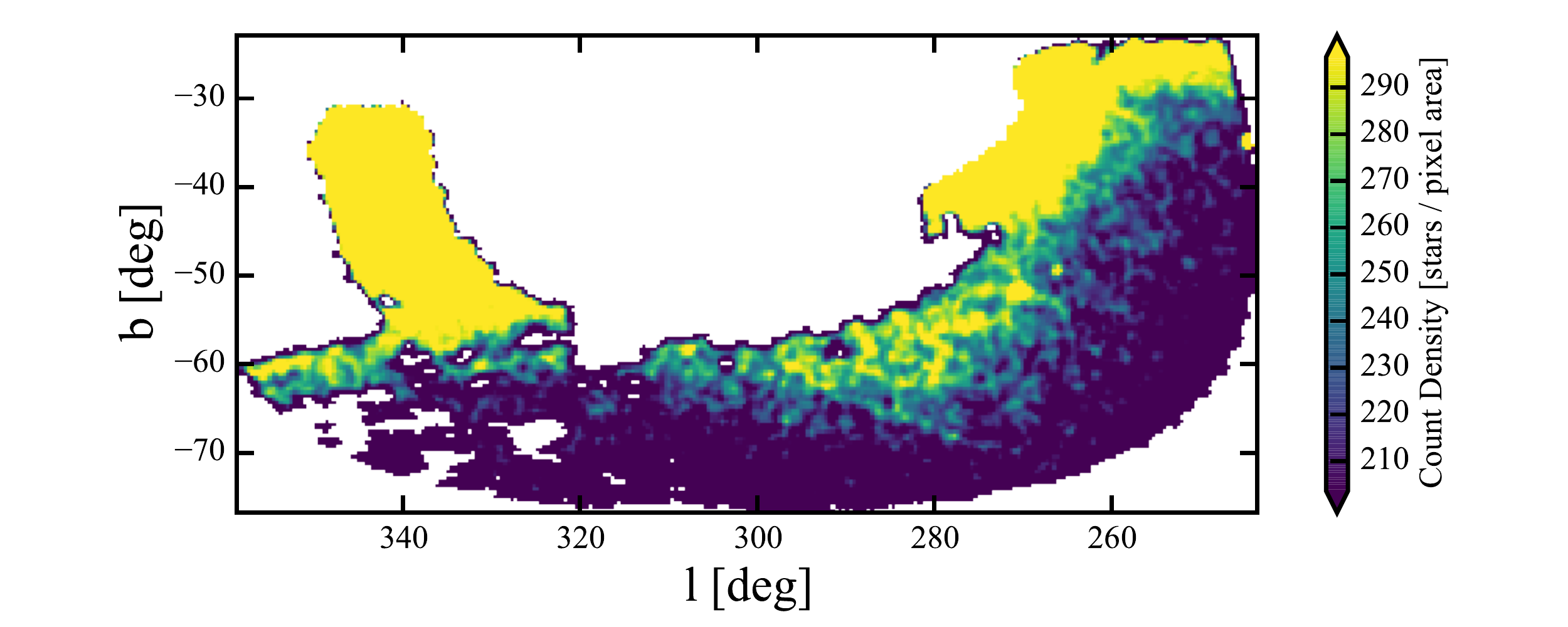}
\plotone{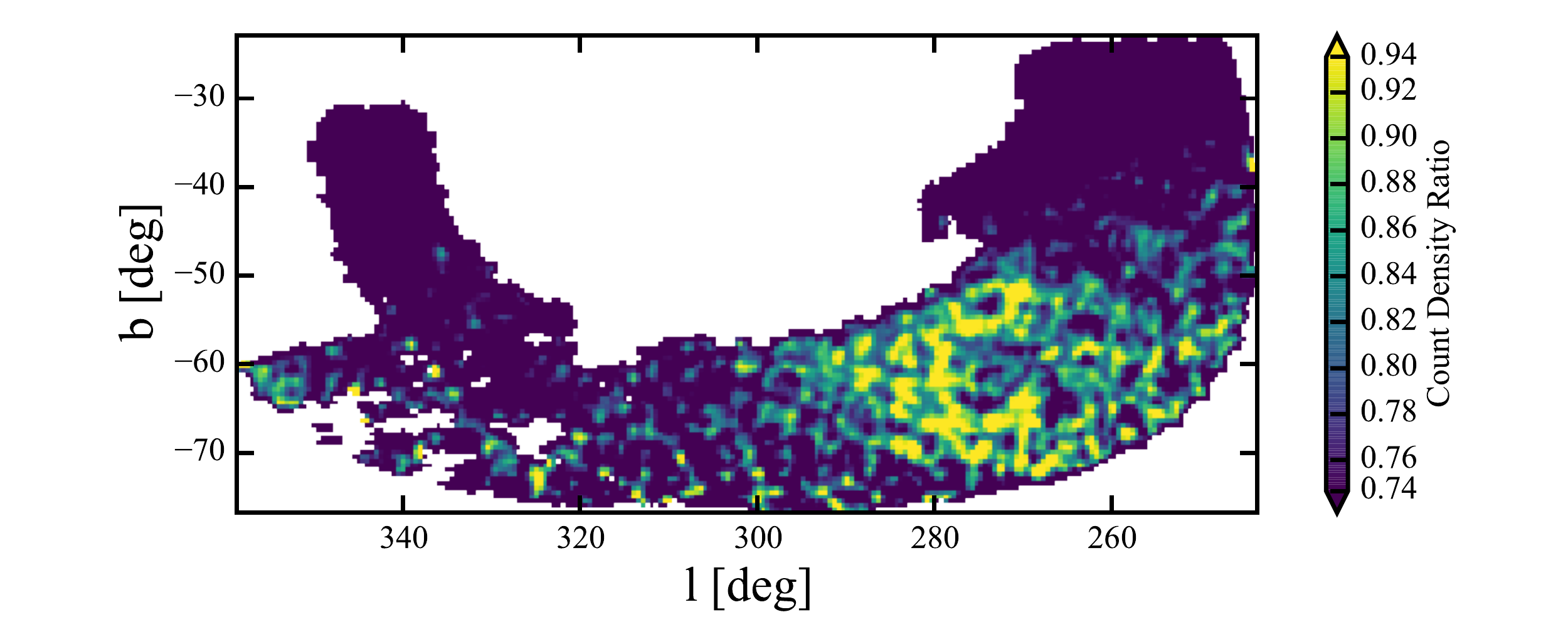}
\plotone{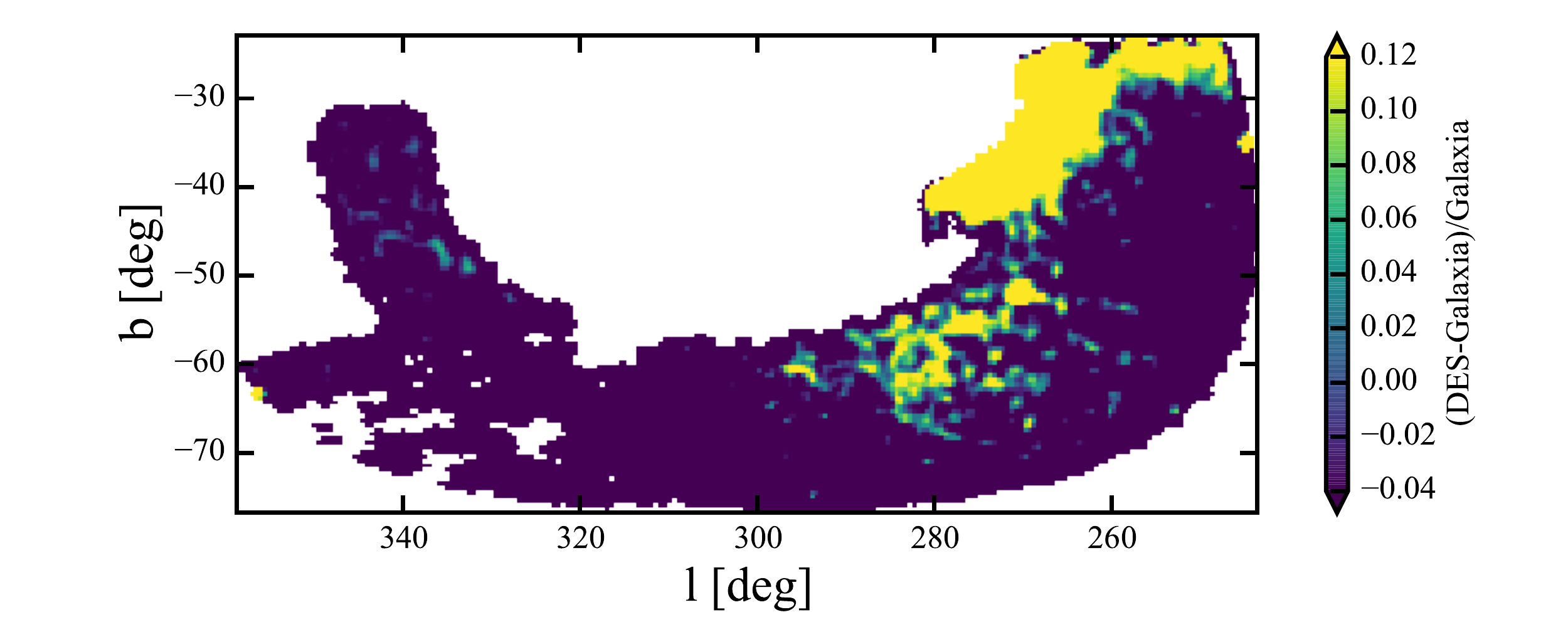}
\caption{\textbf{Top Panel}: The surface density map for the DES SPT field from the Y1A1 data release.  The sample is selected to be stars with $0.2<(g-r)_0<0.4$ and $20.5<g_0<22$.  Each pixel is $\Delta l \times \Delta b = 0.5\degree \times 0.5\degree$. The density for each bin is calculated as the number of stars per pixel area in the unit of $counts/deg^2$, where the pixel area is defined as $\Delta l \times \Delta b \cos b$. The histogram is then smoothed using a 2-d Gaussian kernel with a standard deviation of one bin (i.e., 0.5\degree) in both $l$ and $b$. \textbf{Middle Panel}: The density ratio map for the same area. The density of the stars with $0.2<(g-r)_0<0.4$ and $20.5<g_0<22$ is the numerator and the density of the stars with $0.2<(g-r)_0<0.4$ and $18<g_0<20.5$ and $22<g_0<22.5$ is the denominator. The density is calculated in the same way as the top panel. The histogram is again smoothed using a 2-d Gaussian kernel with a standard deviation of 0.5\degree. \textbf{Bottom Panel}: The fractional difference density map (see text for details) between observed data from DES and mock data taken from a model galaxy constructed using \code{Galaxia}. The same cut, $0.2<(g-r)_0<0.4$ and $20.5<g_0<22$, is applied to both the DES data and the mock data. The histogram is also smoothed using the same Gaussian kernel.  In all three panels, an overdensity around $l\sim285\degree$ and $b\sim-60\degree$ is clearly seen, which is the EriPhe overdensity we refer to in this paper. Note here that the color scale has been chosen to maximize contrast for the EriPhe overdensity. Overflow and underflow pixels are colored according to the triangular extensions of the colorbar.}
\label{fig:density}
\end{figure*}

The upper panel of Figure~\ref{fig:density} shows the density map of stars with $20.5 < g_0 < 22$. Assuming the average F turnoff star with age $\sim$ 12 Gyr has $M_g\sim4.5$,  this magnitude range corresponds to a heliocentric distance of $\sim$15--30 kpc.  The density is calculated as the number of stars per pixel area in unit of $counts/deg^2$. We define the \emph{pixel area} as

\begin{equation}
\delta \Omega = \Delta l \times  \Delta b \cos b
\end{equation}
where $\Delta l$ and  $\Delta b$ are 0.5\degree. The $\cos b$  factor is included so that the actual area of each pixel considered in the density calculation is proportional to the solid angle projected on the sky, even though the Cartesian projection is not an equal area projection.  The map is then smoothed using a 2-D Gaussian kernel with standard deviation of 0.5\degree in both $l$ and $b$. 

\begin{figure*}[th!]
\epsscale{1}
\plotone{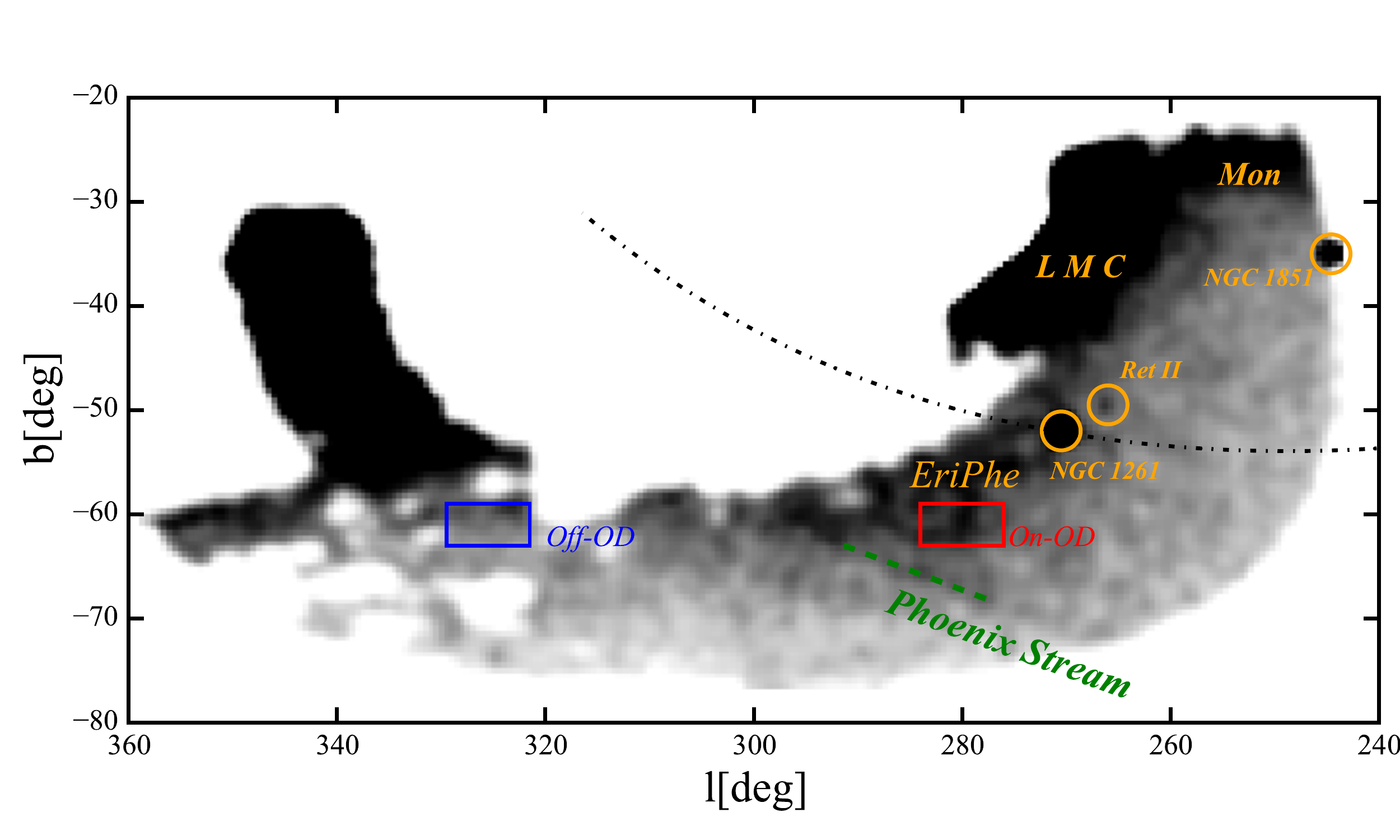}
\caption{
Reproduction of the density map in the top panel of Figure \ref{fig:density}, with previously known Galactic structures identified.  We highlight the overdensities from the LMC, the Monoceros Ring (Mon), and the newly discovered EriPhe. Other interesting objects such as NGC 1261, NGC 1851 and Ret II are labeled in the figure.
The red square box and the blue square box are the on-OD field and off-OD field that used to construct the Hess diagram in Figure \ref{fig:hessdiagram}.  Also plotted are the location of the Phoenix Stream (green dashed line) discovered by~\citet{Balbinotip}, along with the integrated orbit of NGC 1261 (black dashed line) calculated using \code{galpy}~\citep{Bovy2015} and literature values for the proper motions~\citep{Dambis2006} and radial velocity~\citep{Harris2010}. 
}
\label{fig:ill_fig}
\end{figure*}

The two highest density regions in this map are located near $(l,b)\sim(270\degree,-35\degree)$ and $(l,b)\sim(345\degree,-40\degree)$. The former part belongs to the outskirts of the LMC; the latter one is likely the outskirts of the Galactic bulge as it is close to the center of the Galaxy. Apart from those two, we also see an extended overdensity at $270\degree<l<300\degree$  and $-65\degree<b<-55\degree$. We refer to this overdensity as the Eridanus-Phoenix (EriPhe) overdensity. Figure \ref{fig:ill_fig} shows a similar density map to the top panel of Figure \ref{fig:density}, but with all the relevant substructure and fields used in our subsequent analysis explicitly labeled. For instance, it shows that the globular cluster NGC 1261, at $l\sim 270.5\degree$ and $b\sim -52\degree$, is located very near the EriPhe overdensity on the sky, as is the stellar stream in the Phoenix constellation~\citep{Balbinotip}, whose position is also indicated (see more discussion about this in Section \ref{sec:discussion}).

In the middle panel of Figure \ref{fig:density}, we compute the density ratio map. The numerator is the density of stars with $20.5 < g_0 < 22$ and the denominator is the density of stars with $18 < g_0 < 20.5$ and $22 < g_0 < 22.5$. This is essentially the ratio of the number of stars at a distance of $\sim$15--30~kpc, to the number of stars at closer or farther distances.  In this density ratio map, the overdensities from the LMC and the center of the Galaxy in the density map disappear, as the LMC is located at $\sim$ 50 kpc and the stars from the Galactic bulge are located at $<10$~kpc. The EriPhe overdensity stands out much more prominently in the middle panel. It is also worth noting that in this ratio map, the overdensity structure extends as low as b$\sim-70\degree$. However, this extension is not seen in the density map in the top panel. It is possible that EriPhe could extend beyond $b<-65\degree$, but it is hard to determine from the simple binned density map (i.e. top panel), as the background stellar density decreases with increased Galactic latitude. 

We also compared the observational data with mock data generated by the \code{Galaxia} Galactic modeling code~\citep{Sharma2011}. \code{Galaxia} is a code for generating a synthetic catalog of the Galaxy, using either  an analytical input model or one obtained from N-body simulations. The synthetic datasets may be compared to wide-field imaging survey datasets like DES. Here we generated the mock data with a smooth analytical input model.
The same color cut $0.2 < (g-r)_0 < 0.4$ and magnitude cut $20.5 < g_0 < 22$ are applied to the mock data from the \code{Galaxia} model. We then compute the fractional difference of the density map for the observed data from DES minus the \code{Galaxia} model to the \code{Galaxia} model, i.e.
\begin{equation}
\mathrm{fractional\ difference} = \frac{\code{DES} - \code{Galaxia}} {\code{Galaxia}} 
\end{equation}
as shown in the lower panel of Figure \ref{fig:density}. \code{Galaxia} systematically overpredicts the number of stars in the Galaxy relative to DES observations, because 1) the observed data from the DES Y1A1 catalog is not 100\% complete at $g_0\sim22$ for stars and 2) the stellar counts predicted by \code{Galaxia} may also not be accurate at the depth of $g_0\sim22$.
Nevertheless, the EriPhe overdensity stands out from a smooth background. Besides the EriPhe overdensity and the aforementioned LMC outskirts near ($l$, $b$)$\sim$(270\degree, -35\degree), another overdensity near ($l$, $b$)$\sim$(250\degree, -25\degree) is also visible. This overdensity is very likely to be associated with the Monoceros Ring, which is also labeled in Figure \ref{fig:ill_fig}. We leave further discussion of this overdensity to future work.

From the three plots in Figure \ref{fig:density}, we conclude that the EriPhe overdensity  spans at least $270\degree<l<300\degree$  and $-65\degree<b<-55\degree$. The full scope of the structure may extend farther; our current sample is limited by the coverage of the Y1A1 footprint. 

\begin{figure*}[th!]
\epsscale{1}
\center
\plotone{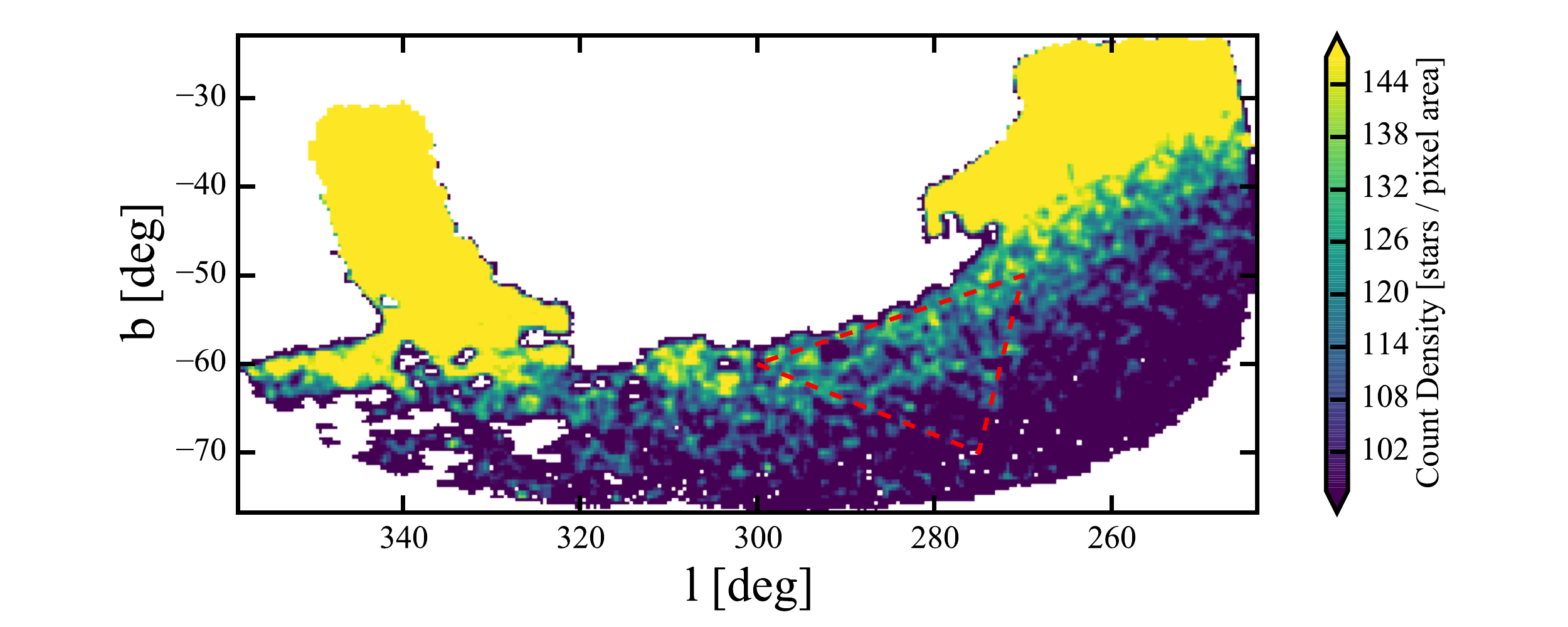}
\plotone{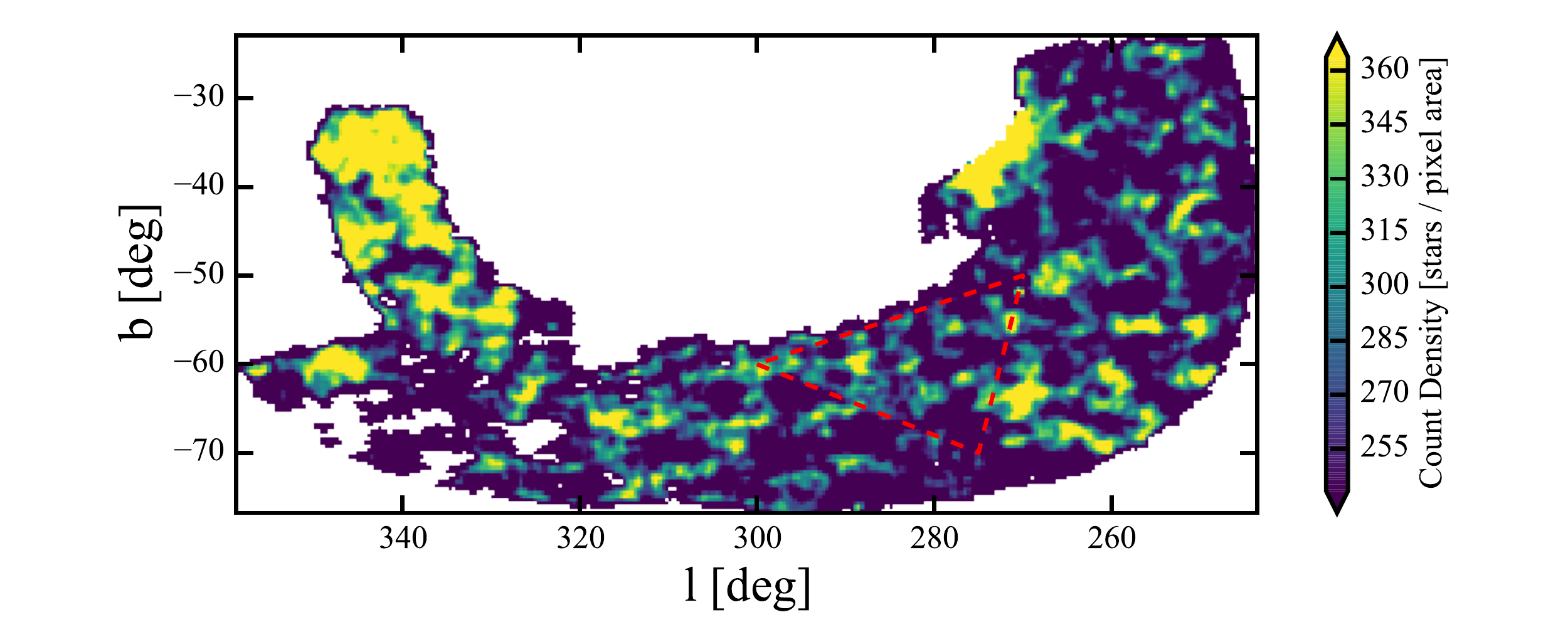}
\plotone{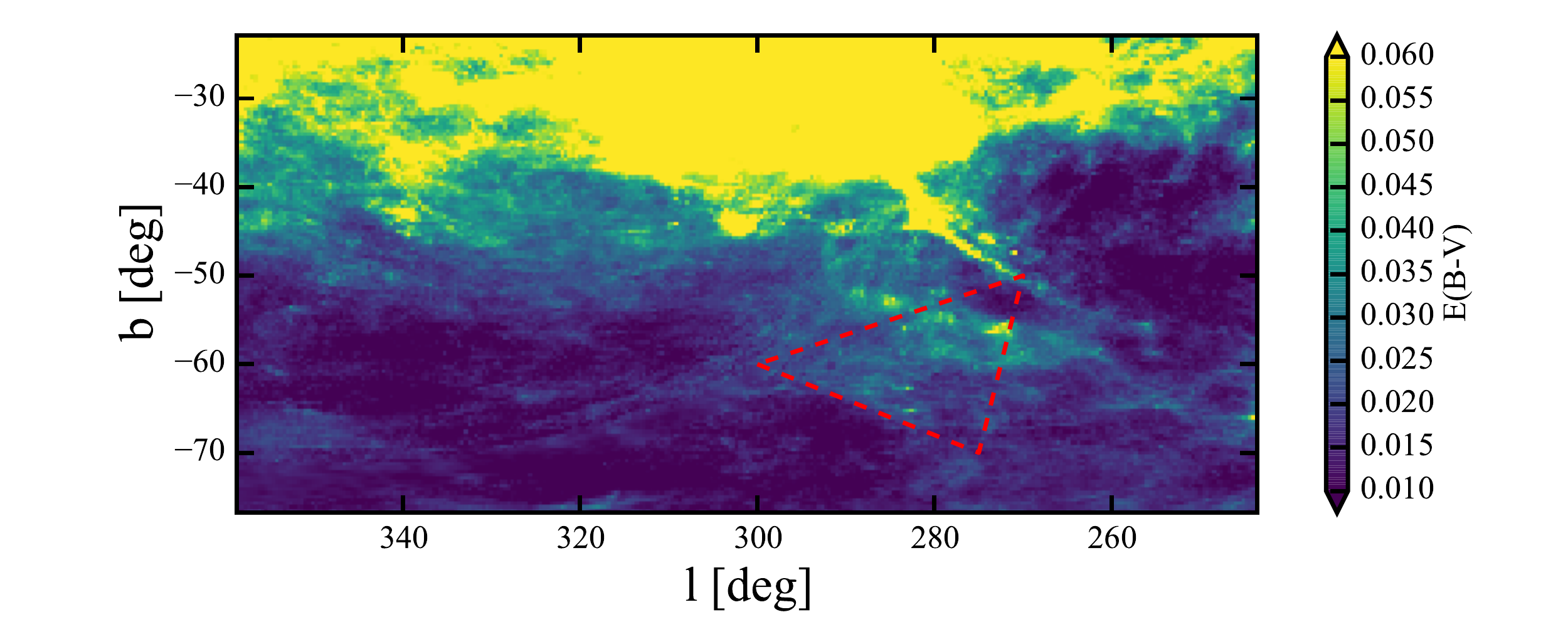}
\caption{
\textbf{Top Panel}: Similar surface density map as the top panel in Figure \ref{fig:density}, but with $0.2<(g-r)_0<0.4$ and $18<g_0<19.5$, to check for overdensities at a difference distance. \textbf{Middle Panel}: Surface density map for non-stellar objects with $0.2<(g-r)_0<0.4$ and $20.5<g_0<22$. The non-stellar object are selected using criteria $|wavg\_spread\_model\_i|>0.003$. \textbf{Bottom Panel}: $E(B-V)$ color excess map from~\citet{Schlegel1998}. In all three panels, the red lines highlight the region of the EriPhe overdensity.
}
\label{fig:density_other}
\end{figure*}

We also perform a series of checks to make sure that this overdensity is not associated with any other systematics. We first check whether the overdensity appears at other heliocentric distance. The upper panel of Figure \ref{fig:density_other} shows the density map of stars with $18 < g_0 < 19.5$ (corresponding to a heliocentric distance of $\sim$5-10~kpc). The red dashed lines highlight the region of EriPhe. It is obvious that there is no overdensity in the same part of the sky at this heliocentric distance. The middle panel of Figure \ref{fig:density_other} shows the density map of non-stellar objects (selected using criteria $|wavg\_spread\_model\_i|>0.003$) with $20.5<g_0<22$ and same color range $0.2<(g-r)_0<0.4$. This panel shows that EriPhe is not associated with the background galaxy distribution. We plot the distribution of the $E(B-V)$ color excess map from \citet{Schlegel1998} in the lower panel of Figure \ref{fig:density_other}. The area enclosed by the red lines has a slightly higher $E(B-V)$ than other areas at the same latitude, by $\Delta E(B-V)\sim0.015$. This difference in the color excess corresponds to a difference of extinction in $g-$band of $\Delta g\sim0.05$~mag, provided that the reddening scaling relation $A_g/E(B-V)=3.237$ from \citet{Schlafly2011}. We therefore investigate the impact of dust obscuration on the EriPhe overdensity. We recalculate the density maps in the same way as in Figure \ref{fig:density} with a uniform dust map correction of $E(B-V)=0.02$, and with no extinction correction. In both cases, the overdensity is still highly significant. We therefore conclude that dust extinction is not a likely explanation of the EriPhe overdensity structure.


\subsection{Hess diagram}
\label{HESS}

We now examine the EriPhe overdensity in color-magnitude space. We build Hess diagrams constructed for the region of the sky that includes the overdensity (hereafter the on-OD field) and for a background region that appears less affected by this structure (hereafter the off-OD field). We construct a Hess diagram using stars having $-0.5 < (g-r)_0 < 1.0$ and $16 < g_0 <  23$.  We select a 10\degree x 5\degree box centered at $(l,b)\sim(280\degree,-61\degree)$ as the on-OD field (red square in Figure \ref{fig:ill_fig}); and another box of the same size centered at $(326\degree,-61\degree)$ as the off-OD field (blue square in Figure \ref{fig:ill_fig}).  The two fields are chosen to lie at the same latitude to minimize any latitude-dependent density changes. The coordinates are selected to minimize gaps in the off-OD field. However, there is still a small gap in the off-OD field, which is about 2\% of the total area. The number of stars is therefore scaled to the sky coverage of the off-OD field.
 
\begin{figure*}[th!]
\epsscale{1}
\centering
\plotone{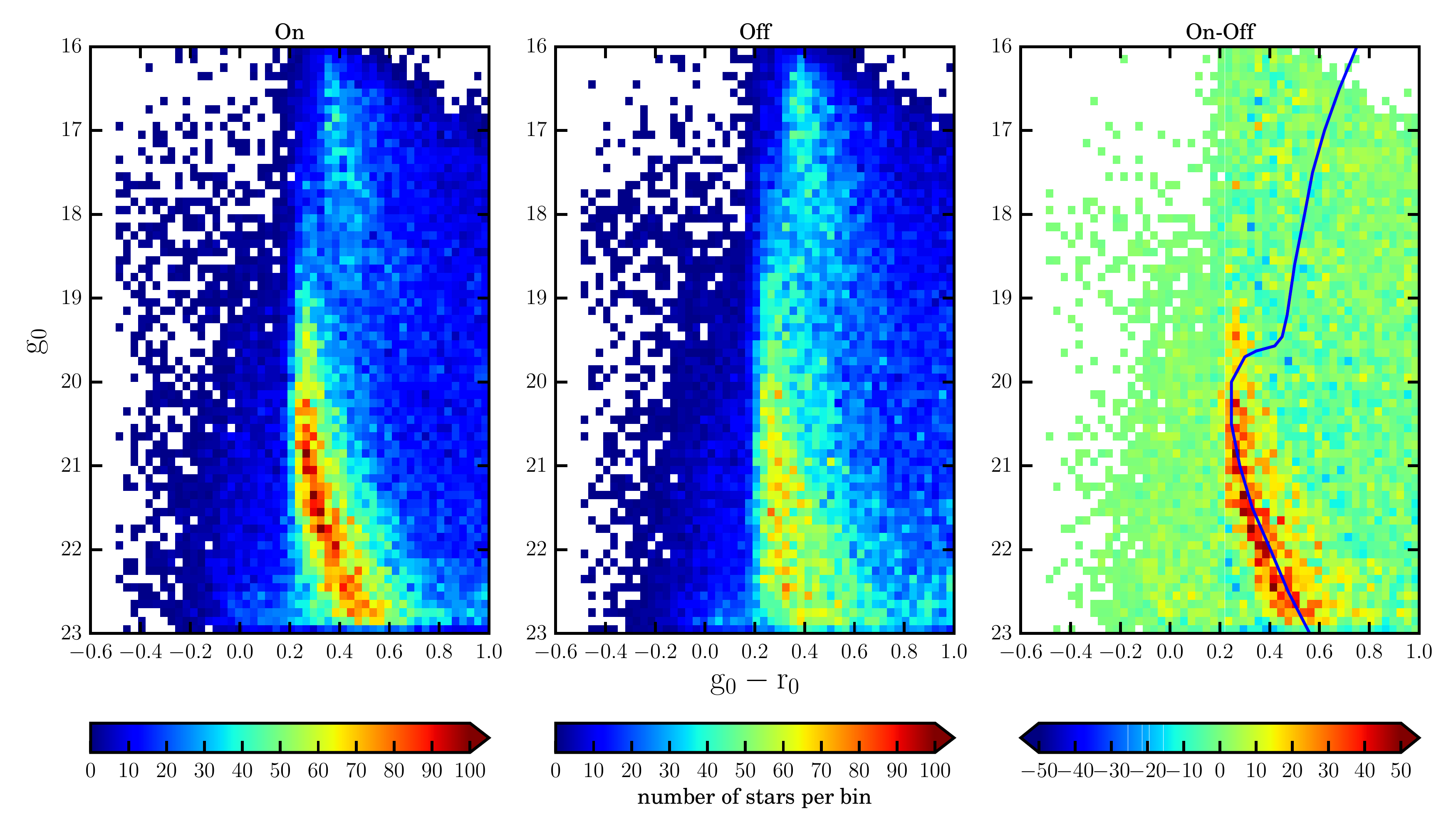}
\caption{Hess diagram for the on-OD field (left panel), a 10\degree x 5\degree box centered at $l=280\degree$, $b=-61\degree$ (i.e. $275\degree<l<285\degree$, $-63.5\degree<l<-58.5\degree$); off-OD field (middle panel), box with same size centered at $l=326\degree$, $b=-61\degree$ (i.e. $321\degree<l<331\degree$, $-63.5\degree<l<-58.5\degree$), and the difference between the two (right panel). The coordinates are selected to minimize gaps in the survey coverage in the off-OD field. However, a small gap in the off-OD field is unavoidable and the off-OD counts are scaled with the sky coverage in this region. The pixel size in each of the three Hess diagrams is $\delta(g-r)_0\times \delta g_0= 0.03\times 0.1$~mag. Also shown as the blue line in the right panel is the ridgeline of NGC 1261 from the DES data.}
\label{fig:hessdiagram}
\end{figure*}

The Hess diagrams for these two fields are shown in the left and middle panels of Figure \ref{fig:hessdiagram}.  The total numbers of stars in the on-OD and off-OD field are 48,710 and 53,525, respectively. The off-OD field has more stars than the on-OD field in total, as it is closer to the Galactic bulge. However, the on-OD field possesses more stars with $0.2 < (g-r)_0 <0.4$ and $20.5< g_0 <  22$ than the off-OD field, 6,409 as compared to 5,429. Considering that the off-OD field has more stars in total, the Poisson significance of the detection is at least $9\sigma$.  We use the same color scale for these two panels to make it apparent that the stars in the on-OD field have a narrower main sequence structure than the off-OD field, even though the on-OD has fewer stars in total. The difference map of the two fields, shown as the right panel in Figure \ref{fig:hessdiagram}, reveals a strong overdensity at $0.2 < (g-r)_0 < 0.5$ and $20.5<g_0<22.5$. Also shown as a blue line in the right panel of Figure \ref{fig:hessdiagram} is the ridgeline of NGC 1261. We obtained the photometry of the stars in NGC 1261 from the DES dataset and derived this empirical isochrone by calculating the median color for each magnitude bin in a step of $\delta g_0=0.2$ for MS stars and $\delta g_0=0.5$ for RGB stars. The isochrone matches the excess in the main sequence in EriPhe, showing that EriPhe has a heliocentric distance that is similar to that of NGC 1261. NGC 1261 has an age of $\sim$~10 Gyr, a metallicity of [Fe/H]~$= -1$ and a heliocentric distance $d=16.3$~kpc~\citep{MarinFranch2009, Forbes2010,Harris2010}. While the NGC 1261 empirical isochrone fits the turn off fairly well, some differences in population are quite possible:
there is some evidence that the age of the overdensity may be younger than that of the cluster as it may have a slightly brighter turnoff, though the red giant branch of the overdensity is too diffuse to be clearly seen in this color-magnitude diagram (see Figure~\ref{fig:hessdiagram}).

We conclude that the Hess diagram analysis demonstrates the existence of a significant population of stars at a common distance around  $l \sim280$\degree and $b\sim-61$\degree, with a magnitude range $20 < g_0 <  22.5$.

\subsection{Surface Brightness and Total Luminosity}
\label{property}

We follow a similar procedure as \citet{Juric2008} to estimate the surface brightness of the EriPhe overdensity.  We sum the fluxes of excess stars in the on-OD field compared to the off-OD field and then divide the total flux by the area of the 10\degree x 5\degree box centered at $b=-61\degree$, which is roughly 25 deg$^2$ or 3.2 x 10$^8 $ arcsec$^2$ projected on the sky.  We only count the fluxes of stars satisfying $0 < (g-r)_0  < 0.7$ and $19 < g_0 < 22.5$ to minimize the contamination from the background. As the off-OD field has a larger total number of stars than the on-OD field, as stated in Section \ref{HESS}, the total sum of the number of counts in the difference map is actually negative -- it is clear to see the overdensity of a main sequence (yellow-to-red coded) and the underdensity elsewhere (blue-coded) in the difference map (right panel of Figure \ref{fig:hessdiagram}). 
We therefore first scaled the number of stars in the off-OD field by a scale factor of 0.80. This is the ratio of the stars with $17 < g_0 <  19$ in two fields, as we assume that the background distribution in this magnitude range is not affected by the overdensity structure. We then calculate the excess stars by using the number of stars in the on-OD field minus the scaled number of stars in the off-OD field. We obtain a value of $\Sigma_r \sim 32.8$~mag~arcsec$^{-2}$. This is about 1.8 mag fainter than the surface brightness of the Sagittarius dwarf Northern stream~\citep{MartinezDelgado2004} and 0.3 mag fainter than that of the Virgo Overdensity~\citep{Juric2008}. This value is effectively a lower limit, because it does not account for stars brighter than $g_0\sim19$ and dimmer than the limiting magnitude ($g_0\sim22.5$). We therefore applied a luminosity correction by adopting the luminosity function and isochrone from~\citet{Dotter2008} with age = 10 Gyr, [Fe/H] $= -1$ and distance modulus $m-M = 16.06$, assuming that the age, metallcity and heliocentric distance of EriPhe is the same as NGC 1261. The surface brightness corrected to the full magnitude range is $\Sigma_r \sim 31.7$~mag~arcsec$^{-2}$. It is worth noting that this calculation assumes that the EriPhe overdensity fully covers the entire 10\degree x 5\degree box. The surface brightness will be underestimated if this 10\degree $\times$ 5\degree box is much larger than the region where an excess of stars related to EriPhe is present. 

We also estimate the total luminosity for EriPhe. First, we assume the surface brightness derived above is uniform across the overdensity structure and equal to $\Sigma_r \sim 31.7$~mag~arcsec$^{-2}$. Second, we assume it spans over $\sim$ 150 deg$^2$ ($270\degree<l<300\degree$  and $-65\degree<b<-55\degree$ is about $\sim$ 150 deg$^2$ projected on the sky). Then, from the surface brightness, we estimate the integrated apparent r-band magnitude to be $m_r = 8.4$. Assuming the heliocentric distance is the same as the globular cluster NGC 1261, we convert the integrated absolute r-band magnitude to be $M_r = -7.7$. Using the relation from~\citet{Juric2008}, we estimate that EriPhe has a total luminosity of $L_r \sim 9 \times 10^4 L\solar$. We acknowledge that this estimate has very large uncertainties. First, the distribution of EriPhe is not uniform and may also be smaller or larger than the assumed 150 deg$^2$. Second, as mentioned earlier, the full scope of EriPhe is limited by the Y1A1 coverage; the structure may extend farther.

We also estimate the total luminosity in an independent way as follows. We first calculate the total number of stars with $0 < (g-r)_0  < 0.7$ and $20.5 < g_0 < 22$ that belong to the EriPhe overdensity by scaling the mock data from \code{Galaxia} so that the number of stars in \code{Galaxia} and the number of stars in the DES data match in the region where the counts are not affected by the EriPhe overdensity, and then summing the counts of the difference between the DES data and the scaled mock data. We get a total of $\sim$21,000 stars in EriPhe with  $0 < (g-r)_0  < 0.7$ and $20.5 < g_0 < 22$. Second, we again adopt the luminosity function and isochrone from~\citet{Dotter2008} and scale the luminosity function so that there are a total of 21,000 stars within the magnitude range $20.5 < g_0 < 22$. Then we sum over the full magnitude range and get a total luminosity of $\sim 11 \times 10^4 L\solar$. The results estimated using these two different methods are roughly consistent. 

\section{Discussion}
\label{sec:discussion}

In the past few years, much theoretical work on the formation of the Milky Way's stellar halo has focused on the tidal stripping and disruption of satellite galaxies~\citep{White1991, Bullock2005, Cooper2010} that are accreted by the Milky Way according to the hierarchical halo formation picture. N-body numerical simulations have predicted that the stellar halo is not a single smooth component, but instead a superposition of many tidal features. Indeed, many halo substructures have been found in the past decade, many of which are believed to be the remnants from past merging events.

In this section we suggest two possible origins of the existence of the EriPhe overdensity formed from tidal stripping and disruption of satellite galaxies.   

\subsection{Connection to NGC 1261 and the Phoenix Stream}

The globular cluster NGC 1261 (highlighted as a circle in Figure \ref{fig:ill_fig}),  at $l\sim 270.5\degree$, $b\sim -52\degree$~\citep{Harris2010}, is located near the edge of the EriPhe overdensity. We note that~\citet{CarballoBello2014} report an unexplained background population around NGC 1261, with similar Galactocentric distance to the cluster, when the region is compared to a Galactic model. They suggested that this population either is disrupted material from the cluster or is an unknown stellar population in the halo. We believe this background is very likely to be part of the EriPhe overdensity. The discovery of EriPhe supports the second suggestion since a disrupting cluster is unlikely to create such a large overdensity on only one side of the cluster.

The EriPhe overdensity is also located very near to the Phoenix Stream found by~\citet{Balbinotip}. Specifically, the Phoenix Stream lies at one edge of the EriPhe overdensity at $(l,b)=(284.4\degree,-66.0\degree)$ to $(l,b)=(294.8\degree,-61.1\degree)$ (highlighted as dashed green lines in Figure \ref{fig:ill_fig}). 
The heliocentric distance of the Phoenix stream is $\sim17.5$~kpc. It is noteworthy that the Phoenix stream, NGC1261, and EriPhe are all located at similar Galactocentric distances ($\sim18$~kpc) and projected locations. 

To determine whether NGC 1261 could be associated with EriPhe and the Phoenix Stream, we compute the orbit of NGC 1261 with \code{galpy}~\citep{Bovy2015} using its proper motions~\citep{Dambis2006} and radial velocity~\citep{Harris2010}. 
The integrated orbit is shown as a black dash-dotted line in Figure \ref{fig:ill_fig}. This orbit roughly aligns with the orientation of the elongated structure of the Phoenix stream, suggesting that these two objects could have shared a Galactic orbit before the presumably more massive structure that resulted in the stream was tidally disrupted along its orbit. Therefore, a possible accretion scenario is that EriPhe is the remains of a dwarf galaxy that initially carried NGC 1261 and the progenitor of the Phoenix stream.  EriPhe and the progenitor of the Phoenix stream could have been tidally disrupted as they were drawn into the Milky Way potential, while NGC 1261 was compact enough to avoid major tidal stripping, or at least none observable with current photometric data. NGC 1261 has an age of $\sim$10 Gyr and a metallicity of [Fe/H]$= -1.0$, suggesting that it is a quite young globular cluster~\citep{MarinFranch2009}.  Many of the Milky Way's other young globular clusters are thought to have originated in dwarf galaxies~\citep{Zinn1993, MarinFranch2009}, and we suggest that the same is true for NGC 1261.

\subsection{A possible polar orbit formed by EriPhe, VOD, and HerAq?}

As discussed in the Introduction, the VOD is a large, diffuse stellar overdensity that extends over 1,000 deg$^2$ in the Northern hemisphere, while HerAq is a similarly diffuse overdensity extending several hundred square degrees.  In comparison with EriPhe at a heliocentric distance of $\sim$16 kpc, the VOD is located at a heliocentric distance of 6-20 kpc, while HerAq has a heliocentric distance of 10-20~kpc.

\begin{figure*}[th!]
\epsscale{1}
\plotone{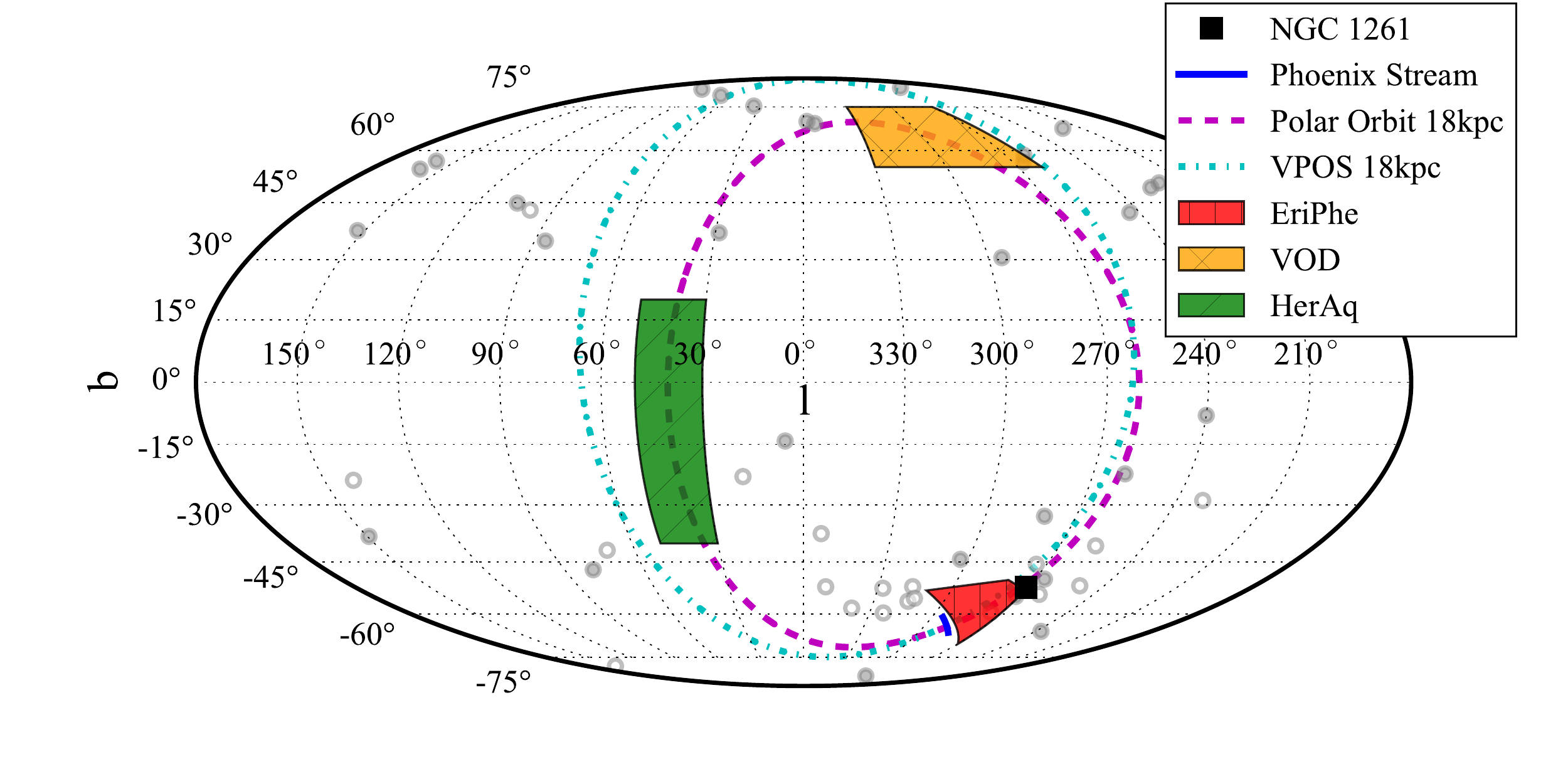}
\caption{Mollweide projection of the known ovendensities discussed in Section \ref{sec:discussion}. A circular polar orbit at a Galactocentric distance of 18 kpc is shown as the purple dashed line. This orbit connects the EriPhe (red), VOD (orange) and HerAq (green), which are highlighted as patches using different colors. As comparison, an orbit at Galactocentric distance of 18 kpc on the VPOS plane using the VPOS+new fit from~\citet{Pawlowski2015} is shown as the cyan dashed-dotted line, together with the confirmed dwarf galaxies in grey filled circles and dwarf galaxy candidates in grey open circles~\citep{McConnachie2012, Bechtol2015,Koposov2015,DrlicaWagner_arXiv, Kim2015a,Kim2015b,Martin2015,Laevens2015}. Also plotted are NGC 1261 (black filled square) and the Phoenix Stream (blue solid line).}
\label{fig:orbit}
\end{figure*}

\citet{Juric2008} proposed a possible explanation for the existence of the VOD as part of a larger ``Polar Ring" structure, forming a ring of overdense regions at the same Galactocentric radii and centered on the Galactic center. However, they were unable to locate a counterpart in the Southern hemisphere by searching the 2MASS catalog for M giants as tracers of the potential ring structure. The discovery of EriPhe makes such a hypothesis more plausible. We carry the Polar Ring suggestion one step further by noting that a circular orbit with radius $\sim18$~kpc from the Galactic center passes through both EriPhe and VOD and intercepts as well the HerAq cloud on the opposite side on the Galactic plane around $l\sim40\degree$ at a heliocentric distance $\sim20$~kpc.  We suggest that the existence and location of these three structures could possibly be explained by a circular Galactic polar orbit. An illustration of the locations and rough spatial extents of the three structures, along with the projected circular polar orbit, is given in Figure \ref{fig:orbit}. This polar plane is also very close to the Vast Polar Structure (VPOS) plane, which is fit to the distribution of Milky Way satellite galaxies~\citep[see, e.g., ][]{Pawlowski2012, Pawlowski2015}. It is therefore possible that EriPhe, VOD, and HerAq are the remnants left behind by a single dwarf satellite near the VPOS plane that has been largely or completely destroyed by the Milky Way's tidal field. 

\citet{Carlin2012} measured the three-dimensional space velocities of stars and showed that the Virgo progenitor was on an eccentric ($e \sim 0.8$) orbit with an inclination of 54\degree. This observational evidence seems to disprove our polar orbit argument. However, \citet{Duffau2014} suggested that VOD appears to be composed of several halo substructures. If the scenario of multiple substructures in the VOD is true, then it is possible that one of them has a polar orbit.

We also note that the circular polar orbit hypothesis would predict the motion of the stars in EriPhe in the opposite directions to the alignment of the Phoenix stream.  Furthermore, the circular orbit cannot explain the discontinuities between VOD, HerAq and EriPhe.
Since the separation of these three overdensities are roughly 120\degree from each other, it is also possible that the disrupted progenitor satellite is on an eccentric orbit with a ``trefoil" shape. These overdensities could be the remnants from the satellite disrupted over several passages, piling up approximately at turning points~\citep[see, e.g., spherical shells in Figure 4 of ][]{Hayashi2003}. Alternatively, these three structures could each have had separate progenitors and have fallen into the Milky Way halo as a group or been accreted along a filament~\citep{Libeskind2005, Zentner2005, Wang2013}. N-body simulations show that tidal stripping is most efficient when the satellites have close passage to the Galactic center and may even lead to total disruption of satellite galaxies~\citep{Bullock2005, Cooper2010}. The resulting diffuse structures are particularly common at small distances from the Galactic center and can be tidal debris from either one single massive progenitor or multiple smaller ones~\citep{Helmi2011}. The validation of either of these hypotheses needs further studies from both observations and simulations.

\section{CONCLUSIONS}
\label{conclusions}
We report the discovery of a Galactic halo stellar overdensity in the Southern hemisphere from the first annual internal release catalog of the first year DES data. This overdensity resides in the constellations of Eridanus and Phoenix, centered around $(l,b) \sim (285\degree,-60\degree)$ and spanning at least 30\degree in longitude and 10\degree in latitude. The full scope of the Eridanus-Phoenix (EriPhe) overdensity may extend farther, but it is limited by the coverage of the first year data. The stellar population in the overdense region has a similar distribution in color-magnitude space as the nearby globular cluster NGC 1261, indicating that the heliocentric distance of EriPhe is about $d\sim16$~kpc and the Galactocentric distance is about $d\sim18$~kpc. We propose two possible scenarios for the creation of the EriPhe overdensity which both involve the tidal stripping and disruption of satellite galaxies. One suggests that EriPhe might be associated with the Phoenix stream and NGC 1261; the other one suggests a possible polar orbit structure formed by EriPhe, VOD, and HerAq stellar overdensities.

Whether or not EriPhe is part of the polar orbit structure, or it is the remnant of a dwarf galaxy with two globular cluster companions (i.e. NGC 1261 and the progenitor of the Phoenix Stream), needs to be tested with future observations. Proper motions measured by Gaia may further inform which hypothesis is favored. Moreover,  spectroscopic follow-up on the Phoenix stream and EriPhe can provide more information on the kinematics and metallicites of these two structures, and improve our understanding of the nature of the EriPhe overdensity. Further study of cloud-like overdensities like EriPhe can also provide insight to probing the gravitational potential of the Milky Way~\citep{Sanderson2013}.

Finally, the second year of DES observations expand the sky coverage of the survey data by adding $\sim$3,000 deg$^2$. The third, fourth, and fifth years of DES observations will increase the depth and uniformity of imaging over the entire 5,000 deg$^2$ survey footprint. Since EriPhe is located near the edge of the footprint of the first year data, a more complete view of EriPhe will be revealed as additional data are acquired.

\acknowledgements{

This paper has gone through internal review by the DES collaboration. We thank
the anonymous referee for comments and suggestions that improved the paper. We also thank Helmut Jerjen and Marcel Pawlowski for providing the original VPOS coordinates. TSL thanks Jonathan Hargis, Steven Boada, Daniel Nagasawa and Katelyn Stringer for very helpful conversations. EBa acknowledges financial support from the European Research Council (ERC-StG-335936, CLUSTERS). This research made use of Astropy, a community-developed core Python package for Astronomy~\citep{Astropy2013}.

Funding for the DES Projects has been provided by the U.S. Department of Energy, the U.S. National Science Foundation, the Ministry of Science and Education of Spain, 
the Science and Technology Facilities Council of the United Kingdom, the Higher Education Funding Council for England, the National Center for Supercomputing 
Applications at the University of Illinois at Urbana-Champaign, the Kavli Institute of Cosmological Physics at the University of Chicago, 
the Center for Cosmology and Astro-Particle Physics at the Ohio State University,
the Mitchell Institute for Fundamental Physics and Astronomy at Texas A\&M University, Financiadora de Estudos e Projetos, 
Funda{\c c}{\~a}o Carlos Chagas Filho de Amparo {\`a} Pesquisa do Estado do Rio de Janeiro, Conselho Nacional de Desenvolvimento Cient{\'i}fico e Tecnol{\'o}gico and 
the Minist{\'e}rio da Ci{\^e}ncia, Tecnologia e Inova{\c c}{\~a}o, the Deutsche Forschungsgemeinschaft and the Collaborating Institutions in the Dark Energy Survey. 

The Collaborating Institutions are Argonne National Laboratory, the University of California at Santa Cruz, the University of Cambridge, Centro de Investigaciones Energ{\'e}ticas, 
Medioambientales y Tecnol{\'o}gicas-Madrid, the University of Chicago, University College London, the DES-Brazil Consortium, the University of Edinburgh, 
the Eidgen{\"o}ssische Technische Hochschule (ETH) Z{\"u}rich, 
Fermi National Accelerator Laboratory, the University of Illinois at Urbana-Champaign, the Institut de Ci{\`e}ncies de l'Espai (IEEC/CSIC), 
the Institut de F{\'i}sica d'Altes Energies, Lawrence Berkeley National Laboratory, the Ludwig-Maximilians Universit{\"a}t M{\"u}nchen and the associated Excellence Cluster Universe, 
the University of Michigan, the National Optical Astronomy Observatory, the University of Nottingham, The Ohio State University, the University of Pennsylvania, the University of Portsmouth, 
SLAC National Accelerator Laboratory, Stanford University, the University of Sussex, and Texas A\&M University.

The DES data management system is supported by the National Science Foundation under Grant Number AST-1138766.
The DES participants from Spanish institutions are partially supported by MINECO under grants AYA2012-39559, ESP2013-48274, FPA2013-47986, and Centro de Excelencia Severo Ochoa SEV-2012-0234.
Research leading to these results has received funding from the European Research Council under the European Union’s Seventh Framework Programme (FP7/2007-2013) including ERC grant agreements 
 240672, 291329, and 306478.

}

\bibliographystyle{apj}
\bibliography{bib}

\begin{thebibliography}{}
\expandafter\ifx\csname natexlab\endcsname\relax\def\natexlab#1{#1}\fi

\bibitem[{Abbott {et~al.}(2005)}]{Abbott2005}
Abbott, T., {et~al.} 2005, arXiv:astro-ph/0510346

\bibitem[{{Astropy Collaboration} {et~al.}(2013){Astropy Collaboration},
  {Robitaille}, {Tollerud}, {Greenfield}, {Droettboom}, {Bray}, {Aldcroft},
  {Davis}, {Ginsburg}, {Price-Whelan}, {Kerzendorf}, {Conley}, {Crighton},
  {Barbary}, {Muna}, {Ferguson}, {Grollier}, {Parikh}, {Nair}, {Unther},
  {Deil}, {Woillez}, {Conseil}, {Kramer}, {Turner}, {Singer}, {Fox}, {Weaver},
  {Zabalza}, {Edwards}, {Azalee Bostroem}, {Burke}, {Casey}, {Crawford},
  {Dencheva}, {Ely}, {Jenness}, {Labrie}, {Lim}, {Pierfederici}, {Pontzen},
  {Ptak}, {Refsdal}, {Servillat}, \& {Streicher}}]{Astropy2013}
{Astropy Collaboration}, {Robitaille}, T.~P., {Tollerud}, E.~J., {et~al.} 2013,
  \aap, 558, A33

\bibitem[{Balbinot {et~al.}(2015)}]{Balbinotip}
Balbinot, {et~al.} 2015, submitted to \apj, arXiv:1509.04283

\bibitem[{{Balbinot} {et~al.}(2013){Balbinot}, {Santiago}, {da Costa}, {Maia},
  {Majewski}, {Nidever}, {Rocha-Pinto}, {Thomas}, {Wechsler}, \&
  {Yanny}}]{Balbinot2013}
{Balbinot}, E., {Santiago}, B.~X., {da Costa}, L., {et~al.} 2013, \apj, 767,
  101

\bibitem[{{Balbinot} {et~al.}(2015){Balbinot}, {Santiago}, {Girardi}, {Pieres},
  {da Costa}, {Maia}, {Gruendl}, {Walker}, {Yanny}, {Drlica-Wagner},
  {Benoit-Levy}, {Abbott}, {Allam}, {Annis}, {Bernstein}, {Bernstein},
  {Bertin}, {Brooks}, {Buckley-Geer}, {Rosell}, {Cunha}, {DePoy}, {Desai},
  {Diehl}, {Doel}, {Estrada}, {Evrard}, {Neto}, {Finley}, {Flaugher},
  {Frieman}, {Gruen}, {Honscheid}, {James}, {Kuehn}, {Kuropatkin}, {Lahav},
  {March}, {Marshall}, {Miller}, {Miquel}, {Ogando}, {Peoples}, {Plazas},
  {Scarpine}, {Schubnell}, {Sevilla-Noarbe}, {Smith}, {Soares-Santos},
  {Suchyta}, {Swanson}, {Tarle}, {Tucker}, {Wechsler}, \&
  {Zuntz}}]{Balbinot2015}
{Balbinot}, E., {Santiago}, B.~X., {Girardi}, L., {et~al.} 2015, \mnras, 449,
  1129

\bibitem[{{Bechtol} {et~al.}(2015){Bechtol}, {Drlica-Wagner}, {Balbinot}, \&
  {Pieres}}]{Bechtol2015}
{Bechtol}, K., {Drlica-Wagner}, A., {Balbinot}, E., \& {Pieres}. 2015, \apj,
  807, 50

\bibitem[{{Belokurov} {et~al.}(2006{\natexlab{a}}){Belokurov}, {Zucker},
  {Evans}, {Wilkinson}, {Irwin}, {Hodgkin}, {Bramich}, {Irwin}, {Gilmore},
  {Willman}, {Vidrih}, {Newberg}, {Wyse}, {Fellhauer}, {Hewett}, {Cole},
  {Bell}, {Beers}, {Rockosi}, {Yanny}, {Grebel}, {Schneider}, {Lupton},
  {Barentine}, {Brewington}, {Brinkmann}, {Harvanek}, {Kleinman}, {Krzesinski},
  {Long}, {Nitta}, {Smith}, \& {Snedden}}]{Belokurov2006a}
{Belokurov}, V., {Zucker}, D.~B., {Evans}, N.~W., {et~al.} 2006{\natexlab{a}},
  \apjl, 647, L111

\bibitem[{{Belokurov} {et~al.}(2006{\natexlab{b}}){Belokurov}, {Zucker},
  {Evans}, {Gilmore}, {Vidrih}, {Bramich}, {Newberg}, {Wyse}, {Irwin},
  {Fellhauer}, {Hewett}, {Walton}, {Wilkinson}, {Cole}, {Yanny}, {Rockosi},
  {Beers}, {Bell}, {Brinkmann}, {Ivezi{\'c}}, \& {Lupton}}]{Belokurov2006b}
---. 2006{\natexlab{b}}, \apjl, 642, L137

\bibitem[{{Belokurov} {et~al.}(2007){Belokurov}, {Evans}, {Bell}, {Irwin},
  {Hewett}, {Koposov}, {Rockosi}, {Gilmore}, {Zucker}, {Fellhauer},
  {Wilkinson}, {Bramich}, {Vidrih}, {Rix}, {Beers}, {Schneider}, {Barentine},
  {Brewington}, {Brinkmann}, {Harvanek}, {Krzesinski}, {Long}, {Pan},
  {Snedden}, {Malanushenko}, \& {Malanushenko}}]{Belokurov2007}
{Belokurov}, V., {Evans}, N.~W., {Bell}, E.~F., {et~al.} 2007, \apjl, 657, L89

\bibitem[{{Bertin}(2011)}]{Bertin2011}
{Bertin}, E. 2011, in Astronomical Society of the Pacific Conference Series,
  Vol. 442, Astronomical Data Analysis Software and Systems XX, ed. I.~N.
  {Evans}, A.~{Accomazzi}, D.~J. {Mink}, \& A.~H. {Rots}, 435

\bibitem[{{Bertin} \& {Arnouts}(1996)}]{Bertin1996}
{Bertin}, E., \& {Arnouts}, S. 1996, \aaps, 117, 393

\bibitem[{{Bonaca} {et~al.}(2012){Bonaca}, {Juri{\'c}}, {Ivezi{\'c}},
  {Bizyaev}, {Brewington}, {Malanushenko}, {Malanushenko}, {Oravetz}, {Pan},
  {Shelden}, {Simmons}, \& {Snedden}}]{Bonaca2012}
{Bonaca}, A., {Juri{\'c}}, M., {Ivezi{\'c}}, {\v Z}., {et~al.} 2012, \aj, 143,
  105

\bibitem[{{Bovy}(2015)}]{Bovy2015}
{Bovy}, J. 2015, \apjs, 216, 29

\bibitem[{{Bullock} \& {Johnston}(2005)}]{Bullock2005}
{Bullock}, J.~S., \& {Johnston}, K.~V. 2005, \apj, 635, 931

\bibitem[{{Carballo-Bello} {et~al.}(2014){Carballo-Bello}, {Sollima},
  {Mart{\'{\i}}nez-Delgado}, {Pila-D{\'{\i}}ez}, {Leaman}, {Fliri},
  {Mu{\~n}oz}, \& {Corral-Santana}}]{CarballoBello2014}
{Carballo-Bello}, J.~A., {Sollima}, A., {Mart{\'{\i}}nez-Delgado}, D., {et~al.}
  2014, \mnras, 445, 2971

\bibitem[{{Carlin} {et~al.}(2012){Carlin}, {Yam}, {Casetti-Dinescu}, {Willett},
  {Newberg}, {Majewski}, \& {Girard}}]{Carlin2012}
{Carlin}, J.~L., {Yam}, W., {Casetti-Dinescu}, D.~I., {et~al.} 2012, \apj, 753,
  145

\bibitem[{{Carlstrom} {et~al.}(2011){Carlstrom}, {Ade}, {Aird}, {Benson},
  {Bleem}, {Busetti}, {Chang}, {Chauvin}, {Cho}, {Crawford}, {Crites}, {Dobbs},
  {Halverson}, {Heimsath}, {Holzapfel}, {Hrubes}, {Joy}, {Keisler}, {Lanting},
  {Lee}, {Leitch}, {Leong}, {Lu}, {Lueker}, {Luong-van}, {McMahon}, {Mehl},
  {Meyer}, {Mohr}, {Montroy}, {Padin}, {Plagge}, {Pryke}, {Ruhl}, {Schaffer},
  {Schwan}, {Shirokoff}, {Spieler}, {Staniszewski}, {Stark}, {Tucker},
  {Vanderlinde}, {Vieira}, \& {Williamson}}]{Carlstrom2011}
{Carlstrom}, J.~E., {Ade}, P.~A.~R., {Aird}, K.~A., {et~al.} 2011, \pasp, 123,
  568

\bibitem[{{Cooper} {et~al.}(2010){Cooper}, {Cole}, {Frenk}, {White}, {Helly},
  {Benson}, {De Lucia}, {Helmi}, {Jenkins}, {Navarro}, {Springel}, \&
  {Wang}}]{Cooper2010}
{Cooper}, A.~P., {Cole}, S., {Frenk}, C.~S., {et~al.} 2010, \mnras, 406, 744

\bibitem[{{Crane} {et~al.}(2003){Crane}, {Majewski}, {Rocha-Pinto},
  {Frinchaboy}, {Skrutskie}, \& {Law}}]{Crane2003}
{Crane}, J.~D., {Majewski}, S.~R., {Rocha-Pinto}, H.~J., {et~al.} 2003, \apjl,
  594, L119

\bibitem[{{Dambis}(2006)}]{Dambis2006}
{Dambis}, A.~K. 2006, Astronomical and Astrophysical Transactions, 25, 185

\bibitem[{{Desai} {et~al.}(2012){Desai}, {Armstrong}, {Mohr}, {Semler}, {Liu},
  {Bertin}, {Allam}, {Barkhouse}, {Bazin}, {Buckley-Geer}, {Cooper}, {Hansen},
  {High}, {Lin}, {Lin}, {Ngeow}, {Rest}, {Song}, {Tucker}, \&
  {Zenteno}}]{Desai2012}
{Desai}, S., {Armstrong}, R., {Mohr}, J.~J., {et~al.} 2012, \apj, 757, 83

\bibitem[{{Diehl} {et~al.}(2014)}]{Diehl2014}
{Diehl}, H.~T., {et~al.} 2014, Proc. SPIE, 9149, 0

\bibitem[{{Dotter} {et~al.}(2008){Dotter}, {Chaboyer}, {Jevremovi{\'c}},
  {Kostov}, {Baron}, \& {Ferguson}}]{Dotter2008}
{Dotter}, A., {Chaboyer}, B., {Jevremovi{\'c}}, D., {et~al.} 2008, \apjs, 178,
  89

\bibitem[{{Drlica-Wagner} {et~al.}(2015){Drlica-Wagner}, {Bechtol}, {Rykoff},
  {Luque}, {Queiroz}, {Mao}, {Wechsler}, {Simon}, {Santiago}, {Yanny},
  {Balbinot}, {Dodelson}, {Fausti Neto}, {James}, {Li}, {Maia}, {Marshall},
  {Pieres}, {Stringer}, {Walker}, {Abbott}, {Abdalla}, {Allam}, {Benoit-Levy},
  {Bernstein}, {Bertin}, {Brooks}, {Buckley-Geer}, {Burke}, {Carnero Rosell},
  {Carrasco Kind}, {Carretero}, {Crocce}, {da Costa}, {Desai}, {Diehl},
  {Dietrich}, {Doel}, {Eifler}, {Evrard}, {Finley}, {Fosalba}, {Frieman},
  {Gaztanaga}, {Gerdes}, {Gruen}, {Gruendl}, {Gutierrez}, {Honscheid}, {Kuehn},
  {Kuropatkin}, {Lahav}, {Martini}, {Miquel}, {Nord}, {Ogando}, {Plazas},
  {Reil}, {Roodman}, {Sako}, {Sanchez}, {Scarpine}, {Schubnell},
  {Sevilla-Noarbe}, {Smith}, {Soares-Santos}, {Sobreira}, {Suchyta}, {Swanson},
  {Tarle}, {Tucker}, {Vikram}, {Wester}, {Zhang}, \&
  {Zuntz}}]{DrlicaWagner_arXiv}
{Drlica-Wagner}, A., {Bechtol}, K., {Rykoff}, E.~S., {et~al.} 2015, ArXiv
  e-prints, arXiv:1508.03622

\bibitem[{{Duffau} {et~al.}(2014){Duffau}, {Vivas}, {Zinn}, {M{\'e}ndez}, \&
  {Ruiz}}]{Duffau2014}
{Duffau}, S., {Vivas}, A.~K., {Zinn}, R., {M{\'e}ndez}, R.~A., \& {Ruiz}, M.~T.
  2014, \aap, 566, A118

\bibitem[{{Duffau} {et~al.}(2006){Duffau}, {Zinn}, {Vivas}, {Carraro},
  {M{\'e}ndez}, {Winnick}, \& {Gallart}}]{Duffau2006}
{Duffau}, S., {Zinn}, R., {Vivas}, A.~K., {et~al.} 2006, \apjl, 636, L97

\bibitem[{{Eggen} {et~al.}(1962){Eggen}, {Lynden-Bell}, \&
  {Sandage}}]{Eggen1962}
{Eggen}, O.~J., {Lynden-Bell}, D., \& {Sandage}, A.~R. 1962, \apj, 136, 748

\bibitem[{{Fitzpatrick}(1999)}]{Fitzpatrick1999}
{Fitzpatrick}, E.~L. 1999, \pasp, 111, 63

\bibitem[{{Flaugher} {et~al.}(2015){Flaugher}, {Diehl}, {Honscheid}, {Abbott},
  {Alvarez}, {Angstadt}, {Annis}, {Antonik}, {Ballester}, {Beaufore},
  {Bernstein}, {Bernstein}, {Bigelow}, {Bonati}, {Boprie}, {Brooks},
  {Buckley-Geer}, {Campa}, {Cardiel-Sas}, {Castander}, {Castilla}, {Cease},
  {Cela-Ruiz}, {Chappa}, {Chi}, {Cooper}, {da Costa}, {Dede}, {Derylo},
  {DePoy}, {de Vicente}, {Doel}, {Drlica-Wagner}, {Eiting}, {Elliott}, {Emes},
  {Estrada}, {Fausti Neto}, {Finley}, {Flores}, {Frieman}, {Gerdes},
  {Gladders}, {Gregory}, {Gutierrez}, {Hao}, {Holland}, {Holm}, {Huffman},
  {Jackson}, {James}, {Jonas}, {Karcher}, {Karliner}, {Kent}, {Kessler},
  {Kozlovsky}, {Kron}, {Kubik}, {Kuehn}, {Kuhlmann}, {Kuk}, {Lahav}, {Lathrop},
  {Lee}, {Levi}, {Lewis}, {Li}, {Mandrichenko}, {Marshall}, {Martinez},
  {Merritt}, {Miquel}, {Munoz}, {Neilsen}, {Nichol}, {Nord}, {Ogando}, {Olsen},
  {Palio}, {Patton}, {Peoples}, {Plazas}, {Rauch}, {Reil}, {Rheault}, {Roe},
  {Rogers}, {Roodman}, {Sanchez}, {Scarpine}, {Schindler}, {Schmidt},
  {Schmitt}, {Schubnell}, {Schultz}, {Schurter}, {Scott}, {Serrano}, {Shaw},
  {Smith}, {Soares-Santos}, {Stefanik}, {Stuermer}, {Suchyta}, {Sypniewski},
  {Tarle}, {Thaler}, {Tighe}, {Tran}, {Tucker}, {Walker}, {Wang}, {Watson},
  {Weaverdyck}, {Wester}, {Woods}, \& {Yanny}}]{flaugher_2015_decam}
{Flaugher}, B., {Diehl}, H.~T., {Honscheid}, K., {et~al.} 2015, submitted to
  AJ, arXiv:1504.02900

\bibitem[{{Font} {et~al.}(2011){Font}, {McCarthy}, {Crain}, {Theuns}, {Schaye},
  {Wiersma}, \& {Dalla Vecchia}}]{Font2011}
{Font}, A.~S., {McCarthy}, I.~G., {Crain}, R.~A., {et~al.} 2011, \mnras, 416,
  2802

\bibitem[{{Forbes} \& {Bridges}(2010)}]{Forbes2010}
{Forbes}, D.~A., \& {Bridges}, T. 2010, \mnras, 404, 1203

\bibitem[{Gruendl {et~al.}(in prep)}]{Gruendlip}
Gruendl, {et~al.} in prep, in prep.

\bibitem[{{Harris}(1996)}]{Harris1996}
{Harris}, W.~E. 1996, \aj, 112, 1487

\bibitem[{{Harris}(2010)}]{Harris2010}
---. 2010, ArXiv e-prints, arXiv:1012.3224

\bibitem[{{Hayashi} {et~al.}(2003){Hayashi}, {Navarro}, {Taylor}, {Stadel}, \&
  {Quinn}}]{Hayashi2003}
{Hayashi}, E., {Navarro}, J.~F., {Taylor}, J.~E., {Stadel}, J., \& {Quinn}, T.
  2003, \apj, 584, 541

\bibitem[{{Helmi} {et~al.}(2011){Helmi}, {Cooper}, {White}, {Cole}, {Frenk}, \&
  {Navarro}}]{Helmi2011}
{Helmi}, A., {Cooper}, A.~P., {White}, S.~D.~M., {et~al.} 2011, \apjl, 733, L7

\bibitem[{{Ivezi{\'c}} {et~al.}(2012){Ivezi{\'c}}, {Beers}, \&
  {Juri{\'c}}}]{Ivezic2012}
{Ivezi{\'c}}, {\v Z}., {Beers}, T.~C., \& {Juri{\'c}}, M. 2012, \araa, 50, 251

\bibitem[{{Johnston} {et~al.}(2008){Johnston}, {Bullock}, {Sharma}, {Font},
  {Robertson}, \& {Leitner}}]{Johnston2008}
{Johnston}, K.~V., {Bullock}, J.~S., {Sharma}, S., {et~al.} 2008, \apj, 689,
  936

\bibitem[{{Johnston} {et~al.}(2012){Johnston}, {Sheffield}, {Majewski},
  {Sharma}, \& {Rocha-Pinto}}]{Johnston2012}
{Johnston}, K.~V., {Sheffield}, A.~A., {Majewski}, S.~R., {Sharma}, S., \&
  {Rocha-Pinto}, H.~J. 2012, \apj, 760, 95

\bibitem[{{Juri{\'c}} {et~al.}(2008){Juri{\'c}}, {Ivezi{\'c}}, {Brooks},
  {Lupton}, {Schlegel}, {Finkbeiner}, {Padmanabhan}, {Bond}, {Sesar},
  {Rockosi}, {Knapp}, {Gunn}, {Sumi}, {Schneider}, {Barentine}, {Brewington},
  {Brinkmann}, {Fukugita}, {Harvanek}, {Kleinman}, {Krzesinski}, {Long},
  {Neilsen}, {Nitta}, {Snedden}, \& {York}}]{Juric2008}
{Juri{\'c}}, M., {Ivezi{\'c}}, {\v Z}., {Brooks}, A., {et~al.} 2008, \apj, 673,
  864

\bibitem[{{Kim} \& {Jerjen}(2015)}]{Kim2015b}
{Kim}, D., \& {Jerjen}, H. 2015, \apjl, 808, L39

\bibitem[{{Kim} {et~al.}(2015){Kim}, {Jerjen}, {Mackey}, {Da Costa}, \&
  {Milone}}]{Kim2015a}
{Kim}, D., {Jerjen}, H., {Mackey}, D., {Da Costa}, G.~S., \& {Milone}, A.~P.
  2015, \apjl, 804, L44

\bibitem[{{Kollmeier} {et~al.}(2009){Kollmeier}, {Gould}, {Shectman},
  {Thompson}, {Preston}, {Simon}, {Crane}, {Ivezi{\'c}}, \&
  {Sesar}}]{Kollmeier2009}
{Kollmeier}, J.~A., {Gould}, A., {Shectman}, S., {et~al.} 2009, \apjl, 705,
  L158

\bibitem[{{Koposov} {et~al.}(2007){Koposov}, {de Jong}, {Belokurov}, {Rix},
  {Zucker}, {Evans}, {Gilmore}, {Irwin}, \& {Bell}}]{Koposov2007}
{Koposov}, S., {de Jong}, J.~T.~A., {Belokurov}, V., {et~al.} 2007, \apj, 669,
  337

\bibitem[{{Koposov} {et~al.}(2015){Koposov}, {Belokurov}, {Torrealba}, \&
  {Evans}}]{Koposov2015}
{Koposov}, S.~E., {Belokurov}, V., {Torrealba}, G., \& {Evans}, N.~W. 2015,
  \apj, 805, 130

\bibitem[{{Laevens} {et~al.}(2015){Laevens}, {Martin}, {Bernard}, {Schlafly},
  {Sesar}, {Rix}, {Bell}, {Ferguson}, {Slater}, {Sweeney}, {Wyse}, {Huxor},
  {Burgett}, {Chambers}, {Draper}, {Hodapp}, {Kaiser}, {Magnier}, {Metcalfe},
  {Tonry}, {Wainscoat}, \& {Waters}}]{Laevens2015}
{Laevens}, B.~P.~M., {Martin}, N.~F., {Bernard}, E.~J., {et~al.} 2015, \apj,
  813, 44

\bibitem[{{Libeskind} {et~al.}(2005){Libeskind}, {Frenk}, {Cole}, {Helly},
  {Jenkins}, {Navarro}, \& {Power}}]{Libeskind2005}
{Libeskind}, N.~I., {Frenk}, C.~S., {Cole}, S., {et~al.} 2005, \mnras, 363, 146

\bibitem[{{Majewski} {et~al.}(2003){Majewski}, {Skrutskie}, {Weinberg}, \&
  {Ostheimer}}]{Majewski2003}
{Majewski}, S.~R., {Skrutskie}, M.~F., {Weinberg}, M.~D., \& {Ostheimer}, J.~C.
  2003, \apj, 599, 1082

\bibitem[{{Mar{\'{\i}}n-Franch} {et~al.}(2009){Mar{\'{\i}}n-Franch},
  {Aparicio}, {Piotto}, {Rosenberg}, {Chaboyer}, {Sarajedini}, {Siegel},
  {Anderson}, {Bedin}, {Dotter}, {Hempel}, {King}, {Majewski}, {Milone},
  {Paust}, \& {Reid}}]{MarinFranch2009}
{Mar{\'{\i}}n-Franch}, A., {Aparicio}, A., {Piotto}, G., {et~al.} 2009, \apj,
  694, 1498

\bibitem[{{Martin} {et~al.}(2004){Martin}, {Ibata}, {Bellazzini}, {Irwin},
  {Lewis}, \& {Dehnen}}]{Martin2004}
{Martin}, N.~F., {Ibata}, R.~A., {Bellazzini}, M., {et~al.} 2004, \mnras, 348,
  12

\bibitem[{{Martin} {et~al.}(2015){Martin}, {Nidever}, {Besla}, {Olsen},
  {Walker}, {Vivas}, {Gruendl}, {Kaleida}, {Mu{\~n}oz}, {Blum}, {Saha}, {Conn},
  {Bell}, {Chu}, {Cioni}, {de Boer}, {Gallart}, {Jin}, {Kunder}, {Majewski},
  {Martinez-Delgado}, {Monachesi}, {Monelli}, {Monteagudo}, {No{\"e}l},
  {Olszewski}, {Stringfellow}, {van der Marel}, \& {Zaritsky}}]{Martin2015}
{Martin}, N.~F., {Nidever}, D.~L., {Besla}, G., {et~al.} 2015, \apjl, 804, L5

\bibitem[{{Mart{\'{\i}}nez-Delgado} {et~al.}(2004){Mart{\'{\i}}nez-Delgado},
  {G{\'o}mez-Flechoso}, {Aparicio}, \& {Carrera}}]{MartinezDelgado2004}
{Mart{\'{\i}}nez-Delgado}, D., {G{\'o}mez-Flechoso}, M.~{\'A}., {Aparicio}, A.,
  \& {Carrera}, R. 2004, \apj, 601, 242

\bibitem[{{Mateo}(1998)}]{Mateo1998}
{Mateo}, M.~L. 1998, \araa, 36, 435

\bibitem[{{McConnachie}(2012)}]{McConnachie2012}
{McConnachie}, A.~W. 2012, \aj, 144, 4

\bibitem[{{Mohr} {et~al.}(2012){Mohr}, {Armstrong}, {Bertin}, {Daues}, {Desai},
  {Gower}, {Gruendl}, {Hanlon}, {Kuropatkin}, {Lin}, {Marriner}, {Petravic},
  {Sevilla}, {Swanson}, {Tomashek}, {Tucker}, \& {Yanny}}]{Mohr2012}
{Mohr}, J.~J., {Armstrong}, R., {Bertin}, E., {et~al.} 2012, Proc. SPIE, 8451,
  84510D

\bibitem[{{Newberg} {et~al.}(2010){Newberg}, {Willett}, {Yanny}, \&
  {Xu}}]{Newberg2010}
{Newberg}, H.~J., {Willett}, B.~A., {Yanny}, B., \& {Xu}, Y. 2010, \apj, 711,
  32

\bibitem[{{Newberg} {et~al.}(2002){Newberg}, {Yanny}, {Rockosi}, {Grebel},
  {Rix}, {Brinkmann}, {Csabai}, {Hennessy}, {Hindsley}, {Ibata}, {Ivezi{\'c}},
  {Lamb}, {Nash}, {Odenkirchen}, {Rave}, {Schneider}, {Smith}, {Stolte}, \&
  {York}}]{Newberg2002}
{Newberg}, H.~J., {Yanny}, B., {Rockosi}, C., {et~al.} 2002, \apj, 569, 245

\bibitem[{{Nie} {et~al.}(2015){Nie}, {Smith}, {Belokurov}, {Fan}, {Fan},
  {Irwin}, {Jiang}, {Jing}, {Koposov}, {Lesser}, {Ma}, {Shen}, {Wang}, {Wu},
  {Zhang}, {Zhou}, {Zhou}, \& {Zou}}]{Nie2015}
{Nie}, J.~D., {Smith}, M.~C., {Belokurov}, V., {et~al.} 2015, ArXiv e-prints,
  arXiv:1508.01272

\bibitem[{{Odenkirchen} {et~al.}(2001){Odenkirchen}, {Grebel}, {Rockosi},
  {Dehnen}, {Ibata}, {Rix}, {Stolte}, {Wolf}, {Anderson}, {Bahcall},
  {Brinkmann}, {Csabai}, {Hennessy}, {Hindsley}, {Ivezi{\'c}}, {Lupton},
  {Munn}, {Pier}, {Stoughton}, \& {York}}]{Odenkirchen2001}
{Odenkirchen}, M., {Grebel}, E.~K., {Rockosi}, C.~M., {et~al.} 2001, \apjl,
  548, L165

\bibitem[{{Pawlowski} {et~al.}(2015){Pawlowski}, {McGaugh}, \&
  {Jerjen}}]{Pawlowski2015}
{Pawlowski}, M.~S., {McGaugh}, S.~S., \& {Jerjen}, H. 2015, ArXiv e-prints,
  arXiv:1505.07465

\bibitem[{{Pawlowski} {et~al.}(2012){Pawlowski}, {Pflamm-Altenburg}, \&
  {Kroupa}}]{Pawlowski2012}
{Pawlowski}, M.~S., {Pflamm-Altenburg}, J., \& {Kroupa}, P. 2012, \mnras, 423,
  1109

\bibitem[{{Pe{\~n}arrubia} {et~al.}(2005){Pe{\~n}arrubia},
  {Mart{\'{\i}}nez-Delgado}, {Rix}, {G{\'o}mez-Flechoso}, {Munn}, {Newberg},
  {Bell}, {Yanny}, {Zucker}, \& {Grebel}}]{Penarrubia2005}
{Pe{\~n}arrubia}, J., {Mart{\'{\i}}nez-Delgado}, D., {Rix}, H.~W., {et~al.}
  2005, \apj, 626, 128

\bibitem[{{Price-Whelan} {et~al.}(2015){Price-Whelan}, {Johnston}, {Sheffield},
  {Laporte}, \& {Sesar}}]{PriceWhelan2015}
{Price-Whelan}, A.~M., {Johnston}, K.~V., {Sheffield}, A.~A., {Laporte},
  C.~F.~P., \& {Sesar}, B. 2015, \mnras, 452, 676

\bibitem[{{Rocha-Pinto} {et~al.}(2004){Rocha-Pinto}, {Majewski}, {Skrutskie},
  {Crane}, \& {Patterson}}]{Rocha2004}
{Rocha-Pinto}, H.~J., {Majewski}, S.~R., {Skrutskie}, M.~F., {Crane}, J.~D., \&
  {Patterson}, R.~J. 2004, \apj, 615, 732

\bibitem[{{Rossetto} {et~al.}(2011){Rossetto}, {Santiago}, {Girardi},
  {Camargo}, {Balbinot}, {da Costa}, {Yanny}, {Maia}, {Makler}, {Ogando},
  {Pellegrini}, {Ramos}, {de Simoni}, {Armstrong}, {Bertin}, {Desai},
  {Kuropatkin}, {Lin}, {Mohr}, \& {Tucker}}]{Rossetto2011}
{Rossetto}, B.~M., {Santiago}, B.~X., {Girardi}, L., {et~al.} 2011, \aj, 141,
  185

\bibitem[{{Sanderson} \& {Helmi}(2013)}]{Sanderson2013}
{Sanderson}, R.~E., \& {Helmi}, A. 2013, \mnras, 435, 378

\bibitem[{{Schlafly} \& {Finkbeiner}(2011)}]{Schlafly2011}
{Schlafly}, E.~F., \& {Finkbeiner}, D.~P. 2011, \apj, 737, 103

\bibitem[{{Schlegel} {et~al.}(1998){Schlegel}, {Finkbeiner}, \&
  {Davis}}]{Schlegel1998}
{Schlegel}, D.~J., {Finkbeiner}, D.~P., \& {Davis}, M. 1998, \apj, 500, 525

\bibitem[{{Searle} \& {Zinn}(1978)}]{Searle1978}
{Searle}, L., \& {Zinn}, R. 1978, \apj, 225, 357

\bibitem[{{Sesar} {et~al.}(2010){Sesar}, {Vivas}, {Duffau}, \&
  {Ivezi{\'c}}}]{Sesar2010}
{Sesar}, B., {Vivas}, A.~K., {Duffau}, S., \& {Ivezi{\'c}}, {\v Z}. 2010, \apj,
  717, 133

\bibitem[{{Sesar} {et~al.}(2007){Sesar}, {Ivezi{\'c}}, {Lupton}, {Juri{\'c}},
  {Gunn}, {Knapp}, {DeLee}, {Smith}, {Miknaitis}, {Lin}, {Tucker}, {Doi},
  {Tanaka}, {Fukugita}, {Holtzman}, {Kent}, {Yanny}, {Schlegel}, {Finkbeiner},
  {Padmanabhan}, {Rockosi}, {Bond}, {Lee}, {Stoughton}, {Jester}, {Harris},
  {Harding}, {Brinkmann}, {Schneider}, {York}, {Richmond}, \& {Vanden
  Berk}}]{Sesar2007}
{Sesar}, B., {Ivezi{\'c}}, {\v Z}., {Lupton}, R.~H., {et~al.} 2007, \aj, 134,
  2236

\bibitem[{{Sevilla} {et~al.}(2011){Sevilla}, {Armstrong}, {Bertin}, {Carlson},
  {Daues}, {Desai}, {Gower}, {Gruendl}, {Hanlon}, {Jarvis}, {Kessler},
  {Kuropatkin}, {Lin}, {Marriner}, {Mohr}, {Petravick}, {Sheldon}, {Swanson},
  {Tomashek}, {Tucker}, {Yang}, {Yanny}, \& {for the DES
  Collaboration}}]{Sevilla2011}
{Sevilla}, I., {Armstrong}, R., {Bertin}, E., {et~al.} 2011, ArXiv e-prints,
  arXiv:1109.6741

\bibitem[{{Sharma} {et~al.}(2011){Sharma}, {Bland-Hawthorn}, {Johnston}, \&
  {Binney}}]{Sharma2011}
{Sharma}, S., {Bland-Hawthorn}, J., {Johnston}, K.~V., \& {Binney}, J. 2011,
  \apj, 730, 3

\bibitem[{{Sharma} {et~al.}(2010){Sharma}, {Johnston}, {Majewski}, {Mu{\~n}oz},
  {Carlberg}, \& {Bullock}}]{Sharma2010}
{Sharma}, S., {Johnston}, K.~V., {Majewski}, S.~R., {et~al.} 2010, \apj, 722,
  750

\bibitem[{{Simion} {et~al.}(2014){Simion}, {Belokurov}, {Irwin}, \&
  {Koposov}}]{Simion2014}
{Simion}, I.~T., {Belokurov}, V., {Irwin}, M., \& {Koposov}, S.~E. 2014,
  \mnras, 440, 161

\bibitem[{{Skrutskie} {et~al.}(2006){Skrutskie}, {Cutri}, {Stiening},
  {Weinberg}, {Schneider}, {Carpenter}, {Beichman}, {Capps}, {Chester},
  {Elias}, {Huchra}, {Liebert}, {Lonsdale}, {Monet}, {Price}, {Seitzer},
  {Jarrett}, {Kirkpatrick}, {Gizis}, {Howard}, {Evans}, {Fowler}, {Fullmer},
  {Hurt}, {Light}, {Kopan}, {Marsh}, {McCallon}, {Tam}, {Van Dyk}, \&
  {Wheelock}}]{Skrutskie2006}
{Skrutskie}, M.~F., {Cutri}, R.~M., {Stiening}, R., {et~al.} 2006, \aj, 131,
  1163

\bibitem[{{Slater} {et~al.}(2014){Slater}, {Bell}, {Schlafly}, {Morganson},
  {Martin}, {Rix}, {Pe{\~n}arrubia}, {Bernard}, {Ferguson}, {Martinez-Delgado},
  {Wyse}, {Burgett}, {Chambers}, {Draper}, {Hodapp}, {Kaiser}, {Magnier},
  {Metcalfe}, {Price}, {Tonry}, {Wainscoat}, \& {Waters}}]{Slater2014}
{Slater}, C.~T., {Bell}, E.~F., {Schlafly}, E.~F., {et~al.} 2014, \apj, 791, 9

\bibitem[{{Sollima} {et~al.}(2011){Sollima}, {Valls-Gabaud},
  {Martinez-Delgado}, {Fliri}, {Pe{\~n}arrubia}, \& {Hoekstra}}]{Sollima2011}
{Sollima}, A., {Valls-Gabaud}, D., {Martinez-Delgado}, D., {et~al.} 2011,
  \apjl, 730, L6

\bibitem[{{Steinmetz} \& {Navarro}(2002)}]{Steinmetz2002}
{Steinmetz}, M., \& {Navarro}, J.~F. 2002, \na, 7, 155

\bibitem[{{Vivas} {et~al.}(2001){Vivas}, {Zinn}, {Andrews}, {Bailyn}, {Baltay},
  {Coppi}, {Ellman}, {Girard}, {Rabinowitz}, {Schaefer}, {Shin}, {Snyder},
  {Sofia}, {van Altena}, {Abad}, {Bongiovanni}, {Brice{\~n}o}, {Bruzual},
  {Della Prugna}, {Herrera}, {Magris}, {Mateu}, {Pacheco}, {S{\'a}nchez},
  {S{\'a}nchez}, {Schenner}, {Stock}, {Vicente}, {Vieira}, {Ferr{\'{\i}}n},
  {Hernandez}, {Gebhard}, {Honeycutt}, {Mufson}, {Musser}, \&
  {Rengstorf}}]{Vivas2001}
{Vivas}, A.~K., {Zinn}, R., {Andrews}, P., {et~al.} 2001, \apjl, 554, L33

\bibitem[{{Walsh} {et~al.}(2007){Walsh}, {Jerjen}, \& {Willman}}]{Walsh2007}
{Walsh}, S.~M., {Jerjen}, H., \& {Willman}, B. 2007, \apjl, 662, L83

\bibitem[{{Wang} {et~al.}(2013){Wang}, {Frenk}, \& {Cooper}}]{Wang2013}
{Wang}, J., {Frenk}, C.~S., \& {Cooper}, A.~P. 2013, \mnras, 429, 1502

\bibitem[{{Watkins} {et~al.}(2009){Watkins}, {Evans}, {Belokurov}, {Smith},
  {Hewett}, {Bramich}, {Gilmore}, {Irwin}, {Vidrih}, {Wyrzykowski}, \&
  {Zucker}}]{Watkins2009}
{Watkins}, L.~L., {Evans}, N.~W., {Belokurov}, V., {et~al.} 2009, \mnras, 398,
  1757

\bibitem[{{White} \& {Frenk}(1991)}]{White1991}
{White}, S.~D.~M., \& {Frenk}, C.~S. 1991, \apj, 379, 52

\bibitem[{{Willman} {et~al.}(2005{\natexlab{a}}){Willman}, {Blanton}, {West},
  {Dalcanton}, {Hogg}, {Schneider}, {Wherry}, {Yanny}, \&
  {Brinkmann}}]{Willman2005a}
{Willman}, B., {Blanton}, M.~R., {West}, A.~A., {et~al.} 2005{\natexlab{a}},
  \aj, 129, 2692

\bibitem[{{Willman} {et~al.}(2005{\natexlab{b}}){Willman}, {Dalcanton},
  {Martinez-Delgado}, {West}, {Blanton}, {Hogg}, {Barentine}, {Brewington},
  {Harvanek}, {Kleinman}, {Krzesinski}, {Long}, {Neilsen}, {Nitta}, \&
  {Snedden}}]{Willman2005b}
{Willman}, B., {Dalcanton}, J.~J., {Martinez-Delgado}, D., {et~al.}
  2005{\natexlab{b}}, \apjl, 626, L85

\bibitem[{{Xu} {et~al.}(2015){Xu}, {Newberg}, {Carlin}, {Liu}, {Deng}, {Li},
  {Sch{\"o}nrich}, \& {Yanny}}]{Xu2015}
{Xu}, Y., {Newberg}, H.~J., {Carlin}, J.~L., {et~al.} 2015, \apj, 801, 105

\bibitem[{{Yanny} {et~al.}(2003){Yanny}, {Newberg}, {Grebel}, {Kent},
  {Odenkirchen}, {Rockosi}, {Schlegel}, {Subbarao}, {Brinkmann}, {Fukugita},
  {Ivezic}, {Lamb}, {Schneider}, \& {York}}]{Yanny2003}
{Yanny}, B., {Newberg}, H.~J., {Grebel}, E.~K., {et~al.} 2003, \apj, 588, 824

\bibitem[{{York} {et~al.}(2000){York}, {Adelman}, {Anderson}, {Anderson},
  {Annis}, {Bahcall}, {Bakken}, {Barkhouser}, {Bastian}, {Berman}, {Boroski},
  {Bracker}, {Briegel}, {Briggs}, {Brinkmann}, {Brunner}, {Burles}, {Carey},
  {Carr}, {Castander}, {Chen}, {Colestock}, {Connolly}, {Crocker}, {Csabai},
  {Czarapata}, {Davis}, {Doi}, {Dombeck}, {Eisenstein}, {Ellman}, {Elms},
  {Evans}, {Fan}, {Federwitz}, {Fiscelli}, {Friedman}, {Frieman}, {Fukugita},
  {Gillespie}, {Gunn}, {Gurbani}, {de Haas}, {Haldeman}, {Harris}, {Hayes},
  {Heckman}, {Hennessy}, {Hindsley}, {Holm}, {Holmgren}, {Huang}, {Hull},
  {Husby}, {Ichikawa}, {Ichikawa}, {Ivezi{\'c}}, {Kent}, {Kim}, {Kinney},
  {Klaene}, {Kleinman}, {Kleinman}, {Knapp}, {Korienek}, {Kron}, {Kunszt},
  {Lamb}, {Lee}, {Leger}, {Limmongkol}, {Lindenmeyer}, {Long}, {Loomis},
  {Loveday}, {Lucinio}, {Lupton}, {MacKinnon}, {Mannery}, {Mantsch}, {Margon},
  {McGehee}, {McKay}, {Meiksin}, {Merelli}, {Monet}, {Munn}, {Narayanan},
  {Nash}, {Neilsen}, {Neswold}, {Newberg}, {Nichol}, {Nicinski}, {Nonino},
  {Okada}, {Okamura}, {Ostriker}, {Owen}, {Pauls}, {Peoples}, {Peterson},
  {Petravick}, {Pier}, {Pope}, {Pordes}, {Prosapio}, {Rechenmacher}, {Quinn},
  {Richards}, {Richmond}, {Rivetta}, {Rockosi}, {Ruthmansdorfer}, {Sandford},
  {Schlegel}, {Schneider}, {Sekiguchi}, {Sergey}, {Shimasaku}, {Siegmund},
  {Smee}, {Smith}, {Snedden}, {Stone}, {Stoughton}, {Strauss}, {Stubbs},
  {SubbaRao}, {Szalay}, {Szapudi}, {Szokoly}, {Thakar}, {Tremonti}, {Tucker},
  {Uomoto}, {Vanden Berk}, {Vogeley}, {Waddell}, {Wang}, {Watanabe},
  {Weinberg}, {Yanny}, {Yasuda}, \& {SDSS Collaboration}}]{York2000}
{York}, D.~G., {Adelman}, J., {Anderson}, Jr., J.~E., {et~al.} 2000, \aj, 120,
  1579

\bibitem[{{Zentner} {et~al.}(2005){Zentner}, {Berlind}, {Bullock}, {Kravtsov},
  \& {Wechsler}}]{Zentner2005}
{Zentner}, A.~R., {Berlind}, A.~A., {Bullock}, J.~S., {Kravtsov}, A.~V., \&
  {Wechsler}, R.~H. 2005, \apj, 624, 505

\bibitem[{{Zinn}(1993)}]{Zinn1993}
{Zinn}, R. 1993, in Astronomical Society of the Pacific Conference Series,
  Vol.~48, The Globular Cluster-Galaxy Connection, ed. G.~H. {Smith} \& J.~P.
  {Brodie}, 302

\bibitem[{{Zucker} {et~al.}(2006){Zucker}, {Belokurov}, {Evans}, {Wilkinson},
  {Irwin}, {Sivarani}, {Hodgkin}, {Bramich}, {Irwin}, {Gilmore}, {Willman},
  {Vidrih}, {Fellhauer}, {Hewett}, {Beers}, {Bell}, {Grebel}, {Schneider},
  {Newberg}, {Wyse}, {Rockosi}, {Yanny}, {Lupton}, {Smith}, {Barentine},
  {Brewington}, {Brinkmann}, {Harvanek}, {Kleinman}, {Krzesinski}, {Long},
  {Nitta}, \& {Snedden}}]{Zucker2006}
{Zucker}, D.~B., {Belokurov}, V., {Evans}, N.~W., {et~al.} 2006, \apjl, 643,
  L103

\end{thebibliography}

\end{document}